\documentclass[11pt]{article}

\usepackage{graphicx}
\usepackage{graphicx,psfrag}

\def \V {\vskip 0.5cm }

\def \v {\vskip 0.2cm}

\def \E {{\bf \, E}}

\def \CQ {{\cal Q}}

\def \T {{\hbox{\,Tr\,}}}

\def \I {{\hbox{ In}}}
\def \O {\hbox{ Out}}

\def \CV {{\cal V}}
\def \CE {{\cal E}}
\def \CC {{\cal C}}

\def \CU {{\cal U}}
\def \CN {{\cal N}}
\def \CT {{\cal T}}

\def \CW {{\cal W}}
\def \CD {{\cal D}}
\def \CI {{\cal I}}
\def\CL {{\cal L}}
\def \CQ {{\cal Q}}

\def \CA {{\cal A}}

\def \D {\Delta}

\def \a {\alpha}

\def \b {\beta}

\def \g {\gamma}

\def \s {\sigma}

\def \t {\tau}

\def \d {\delta}

\def \U {\Xi}
\def \u {\xi}

\begin{document}

\title{EVEN   WALKS AND  ESTIMATES OF HIGH MOMENTS OF LARGE WIGNER
RANDOM MATRICES}

\author{O. Khorunzhiy\footnote{Universit\'e de Versailles - Saint-Quentin, Versailles, FRANCE;
{\it e-mail:} khorunjy@math.uvsq.fr} \   and V.
Vengerovsky\footnote{Mathematical Division, B. Verkin Institute
for Low Temperature Physics, Kharkov, UKRAINE}}

\maketitle

\begin{abstract}
We revisit the problem of estimates of  the moments $m_{2s}^{(n)}=
\E \{\T A_n^{2s}\}$ of  random $n\times n$ matrices of the Wigner
ensemble by  using the approach elaborated by Ya. Sinai and
\mbox{A. Soshnikov} and further developed  by A. Ruzmaikina. We
continue to investigate the structure of closed even walks
$w_{2s}$ and their graphs $g(w_{2s})$ that arise in these studies.
One of the key problems here is related with  the graphs
$g(w_{2s})$ that have at least one vertex $\b$ that is the tail of
a large number of edges. This situation occurs when the
corresponding Dyck path (or equivalently, the plane rooted tree)
has a vertex of large degree; in the opposite case this can happen
when the self-intersection degree of $\b$ is large. We show that
there exists one more possibility; it is given by the case when
$w_{2s}$ has a large number of open instants of
self-intersections, or more precisely, a large total number of the
instants of broken tree structure. Basing on this observation, we
modify the technique mentioned above and prove the estimates of
the moments $m_{2s_n}^{(n)}$ in the limit $s_n,n\to \infty$ when
$s_n=O(n^{2/3})$.

\end{abstract}

\section{Introduction}

Random matrices of infinite dimensions represent a very rich and
interesting subject of studies that relates various branches of
mathematical physics, analysis, combinatorics and many others.

The spectral theory of large random matrices started half a
century ago by \mbox{E. Wigner} is still a source of interesting
and  challenging problems. An important part of these problems
concerns the universality conjecture for the local spectral
properties of ensembles of large
 real symmetric (or hermitian) matrices  (see e.g. the
monograph \cite{M}). One of the examples of such ensembles is
given by the Wigner ensemble of $n\times n$ real symmetric random
matrices
 of the form
$$
(A_n)_{ij} = {1\over \sqrt n} a_{ij}, \eqno (1.1)
$$
where $\{ a_{ij}, 1\le i\le j \le  n\}$ are jointly independent
random variables such that the following conditions are verified
$$
\E \{a_{ij}\} = 0, \quad {\hbox {and}} \quad \E \{ a_{ij} ^2\} =
{1}, \eqno (1.2)
$$
where $\E\{\cdot \}$ denotes the mathematical expectation. E.
Wigner proved that the \mbox{normalized} moments of $A_n$ converge
in the limit $n\to\infty$
$$
\lim_{n\to\infty} {1\over n} \E \left\{\T A_n^{p}\right\} =
\cases{ {(2s)!\over  s!\, (s+1)!}, & if $p=2s$,\cr 0, & if
$p=2s+1$\cr}
 \eqno (1.3)
$$
under conditions that  all moments of all $a_{ij}$ exist and the
probability distribution of random variables $a_{ij}$ is symmetric
\cite{W1,W2}.

To study the moments (1.3), E. Wigner  has interpreted  the trace
of the product
$$
\E\left\{ \T A_n^p\right\} =  \sum_{i_0, i_1, \dots, i_p =1}^n \E
\left\{ A_{i_0,i_1}\cdots A_{i_{p-1},i_0} \right\}
 \eqno(1.4)
$$
as the weighted sum over all possible sequences $I_p = (i_0, i_1,
\dots, i_{p-1},i_0)$ \cite{W1}. The set of these sequences can be
separated into classes of equivalence that in the case of even
$p=2s$ can be labelled by simple non-negative  walks of $2s$ steps
$\theta_{2s}$ that start and end at zero. These walks are known as
the Dyck paths and the Catalan number $C^{(s)} = (2s)!/s! (s+1)!$
standing in (1.3) represents the number of all Dyck paths with
$2s$ steps. Obviously, $\E\{\T A_n^{2s+1}\}=0$ due to the
symmetric distribution of random variables $a_{ij}$ and it is
sufficient to consider the even moments   $m_{2s}^{(n)}= \E \{ \T
A_n^{2s}\}$ only.

The method proposed by E. Wigner has been used as the starting
point in the studies of the
 asymptotic behavior of variables
  $m_{2s}^{(n)}$
   in the limit when $s$ goes to infinity at the
same time as $n$ does \cite{BY,FK,G}. In particular, it was shown
in the beginning of 80-s that the estimate
$$
m_{2s}^{(n)} \le {n (2s)!\over s!\, (s+1)!}\, ( 1 +o(1))
 \quad
 {\hbox{ as}}\quad s,n\to\infty
 \eqno (1.5)
$$
is true when $s= s_n = o(n^{1/6})$ as $n\to\infty$ provided random
variables $a_{ij} $ are bounded with probability 1 and their
probability distribution is symmetric \cite{FK}. The method of
\cite{FK} uses certain encoding of the paths $I_{2s}$ additionally
to the Dyck paths representation. More precisely, the graph theory
approach has been developed in \cite{FK}, where, in particular,
the Dyck paths $\theta_{2s}$ are considered as the canonical runs
over the plane rooted trees $\CT_s= \CT(\theta_{2s})$ of $s+1$
vertices.

Regarding the problem of estimates of high moments of large random
matrices, it is  important to determine the maximally possible
rate of $s_n$ such that (1.5) is still valid. Here a breakthrough
step was made in  paper \cite{SS1}, where (1.5) was shown to be
true for  all $s_n = o(n^{1/2})$, $n\to\infty$ in the case when
the moments of random variables $a_{ij}$ are of the sub-Gaussian
form and the probability distribution of $a_{ij}$ is symmetric.
Also the Central Limit Theorem for the random variable
 $4^{-s}\left(m_{2s}^{(n)} - \E\{ m_{2s}^{(n)}\}\right)$
  was proved in this limit. It was shown that the limiting
expressions for the corresponding correlation functions do not
depend on the particular values of the moments of $a_{ij}$. This
was a first step toward the proof of   the universality conjecture
for the Wigner ensembles. It should be stressed that in these
studies the estimate (1.5) represents a key result that is crucial
for the proof of the Central Limit Theorem.

The proof of \cite{SS1} is based on the Wigner's approach added by
 an  important notion of the
self-intersection of the sequence $I_{2s}$; in \cite{SS1} these
sequences are called the paths of $2s$ steps. In certain sense,
the self-intersections of the path from the class $\theta_{2s}$
correspond to gluing  of the vertices of the corresponding tree
$\CT(\theta_{2s})$. Then the self-intersection degree of a vertex
can be determined as a number of arrivals to this vertex by the
edges of this tree. In \cite{SS1}, the set of all paths has been
separated into the classes of equivalence according to the number
of vertices of self-intersections and the self-intersection
degrees of these vertices. This gives the tools to control the
number of paths in the limit $s,n\to\infty$.

This method was modified in the subsequent paper \cite{SS2} to
prove that (1.5) is valid in the limit \mbox{$s_n, n\to\infty$,}
\mbox{$s_n = o(n^{2/3})$}. Here   the family of self-intersections
has been further classified and the notion of the  open vertex of
two-fold self-intersection has been considered for the first time.
The next in turn step has been made in \cite{S} to show that an
estimate of the form (1.5) is true  in the limit
\mbox{$n\to\infty$,} $s_n = O(n^{2/3})$ under the same conditions
on the probability distribution of $a_{ij}$ as in \cite{SS1}.

The technique  of \cite{SS1,SS2,S} was further developed in paper
\cite{R}, where the case of arbitrary distributed random variables
$a_{ij}$ with symmetric law of polynomial decay was considered. It
was indicated in \cite{R} that certain estimate of \cite{S}  was
not established in the full extent and a way was proposed to
complete the proof. However, the way proposed in \cite{R} is
partly correct  and the proof of corresponding estimate  suffers
in its own turn from a serious gap.

Let us briefly explain the problem. Following \cite{SS2,S},   it
was assumed in \cite{R} that for the paths with typical
$\theta_{2s}$ the presence of vertices with large number of steps
"out" is possible only when there is a sufficiently large number
of steps "in"; here the steps "in" and "out" are  meant to be the
steps that correspond to the ascending steps of $\theta_{2s}$.
However, this is not the case. The walk can arrive at a given
vertex $\b$ by the steps that correspond to the descend parts of
the Dyck path $\theta_{2s}$ and bring to $\b$ the edges that do
not belong the ascending steps "in". We  call these edges the
imported ones and refer to the corresponding arriving steps as to
the imported cells. At the end of this paper, we present examples
of paths with a number of imported cells and edges.

We see that the description of the family of paths $\{I_{2s}\}$ of
(1.4) should be improved. This is the main subject of the present
work. We will see that the studies require a number of
generalizations of the notions  introduced in \cite{SS2}. In
particular, the notion of the open vertex of two-fold
self-intersection should be generalized for the case of
self-intersections of any degree. Also the further  analysis of
the open instants of self-intersections leads us to the notion of
the instant of broken tree structure. It plays even more important
role with respect to the estimates of the moments $m_{2s}^{(n)}$
than that played by the open vertices of self-intersections.
Clearly, these new circumstances require essential modifications
of the technique proposed by \cite{R,S,SS2}.

The paper is organized as follows. In Section 2 we repeat some of
the definitions of notions  introduced \cite{SS1,SS2,S} and
\cite{R} and formulate  generalizations of them we need. Then we
present our main result about the primary and imported  cells.
Namely, we prove that the number of imported cells at given vertex
$\b$ is determined by the self-intersection degree of $\b$ and by
the total number of broken tree instants of the path. We show that
the number of the instants of broken tree structure is bounded by
the number of open arrival instants and this finally gives us a
tool to control the number of paths that have vertices of large
degree.

The descriptive part given by Section 2 is followed by Section 3,
where the principles of construction of the set of walks and
corresponding estimates are described. Here we mainly follow the
lines of \cite{SS2,S} with necessary modifications and
specifications. In Section 4 we prove estimates of the form (1.5)
of the averaged traces $\E \{ \T (A_n)^{2s_n}\}$. Our main result
concern the case when the matrix entries are given by bounded
random variables $a_{ij}$.

The reason for this restriction is two-fold. First, this helps us
not to overload the paper and to present clearer the key points of
the technique used. From another side, the description of classes
of paths and walks in terms of vertices of self-intersections does
not fit very well the problem of counting the number of multiple
edges in the graphs of these paths and walks. As a result, the
technique described can be applied to wider classes of random
variables $a_{ij}$ while the optimal conditions are far to be
reached. At the end of the paper we prove one of the possible
results in this direction, where we require that a finite number
of moments exist, $\E \vert a_{ij}\vert^q<\infty$, $q\le q_0$.

 Finally, in Section 5 we consider several examples of the walks
with primary and imported cells and show why the estimates of
\cite{R,S} do not work in the corresponding cases.

\section{Even closed walks}

As we mentioned above, it is natural to consider  (1.4) as  the
sum of weights
$$
\E \left\{ \T A_n^{2s}\right\} = {1\over n^s}
 \sum_{i_0,i_1,\dots,i_{2s-1}=1}^{n}
  \E \left\{ a_{i_0, i_1}\,   \cdots \,
a_{i_{2s-1}, i_0}\right\} = {1\over n^s}
 \sum_{I_{2s}\in {\cal I}_{2s}(n)} \CQ(I_{2s}),
 \eqno (2.1)
$$
where the sequence $I_{2s} $ can be regarded a  trajectory of $2s$
steps
$$
I_{2s} = (i_0, i_1, \dots, i_{2s-1}, i_0), \quad i_l\in \{1,
\dots, n\}
 \eqno (2.2)
$$
and  ${\cal I}_{2s}(n)$ denotes  the set of all such trajectories;
the weight $\CQ(I_{2s})$ is given by the average of the product of
corresponding random variables $a_{ij}$. Here and below we omit
subscripts in $s_n$ when they are not necessary. In papers
\cite{SS1,SS2,S} the sequences $I_{2s}$ are referred as to the
{\it paths}, so we keep this terminology in the present paper.

In the present  section we study  the paths $I_{2s}$ and their
graphs and describe partitions of the set $\CI_{2s}(n)$ into the
classes of equivalence according to the self-intersection
properties of $I_{2s}$. To do this, we  give necessary definitions
based on those of \cite{R,SS1,SS2,S} and consider their
generalizations. Then  we prove our main technical result about
the paths with primary and imported cells.

\subsection{Paths, walks and graphs of walks}

In (2.2), it is convenient to  consider the subscripts of $i_l$ as
the instants of the discrete time. Introducing variable  $0\le
t\le 2s$, we write that $I_{2s}(t) = i_t$. We will also say that
the couple $(t-1,t)$ with $1\le t\le 2s$ represents the  step
number $t$ of the path $I_{2s}$.

We determine the set of vertices visited by the path $I_{2s}$ up to the instant $t$
$$
\CU(I_{2s}; t) = \left\{ I_{2s}(t'), \ 0\le t'\le t\right\}
$$
and denote by $\vert \CU(I_{2s};t)\vert$ its cardinality.

Regarding  a particular path $I_{2s}$, one can introduce
corresponding {\it closed walk} $w(I_{2s};t)$, $0\le t\le 2s$ that
is given by a sequence of $2s$ labels (say, numbers from
$(1,\dots,n)$ or letters).  Also we can determine the {\it minimal
closed  walk}  $w_{min}(t)= w_{min}(I_{2s};t)$ constructed from
$I_{2s}$ by the following recurrence rules: \v 1) at the origin of
time, $w_{min}(0) =1 $; \v 2) if $I_{2s}(t+1)\notin
\CU(I_{2s};t)$, then $w_{min}(t+1)= \vert \CU(I_{2s};t)\vert +1$;

\hskip 0.45cm  if there exists such $t'\le t$ that $I_{2s}(t+1)=
I_{2s}(t')$, then $w_{min}(t+1) = w_{min}(t')$.

\v One can interpret  $w_{min}(t)$ as a path of $2s$ steps, where
the number of each new label is given by the number of different
labels used before increased by one. The following sequences
$$
I_8 = (5,2,1,5,7,3,1,5), \quad w_{min}(I_8) = (1,2,3,1,4,5,3,1).
$$
give an example of  the closed path $I_{2s}$ with $2s = 8$  and of
the  corresponding minimal walk $w_{min}(I_{8})$.  The set of all
possible paths ${\cal I}_{2s}^{(n)}$ can be separated into the
classes of equivalence ${\cal C}(w_{min})$ labelled by the minimal
walks $w_{min}$. We say that $I_{2s}$ and $I'_{2s}$ are
equivalent, $I_{2s}\sim I'_{2s}$ when  $w_{min}(I_{2s})
=w_{min}(I'_{2s})$. In what follows, we consider the minimal walks
only, so we omit the subscript $min$ in $w_{min}$ and write simply
that $w_{2s} = w(I_{2s})$.

Regarding  a minimal  closed walk of $2s$ steps $w_{2s}$, one can
consider the {\it graph of the walk}
 $g_{2s} =  g(w_{2s})$
  with the set of vertices $\CV(g_{2s})$
 and the set of oriented  edges $\CE(g_{2s})$; sometimes we will write that
 $g_{2s}=(\CV,\CE)$.

It is natural to accept equality $\CV(g_{2s})= \CU(w_{2s};2s)$. We
denote the vertices $g(w_{2s})$ with the help of the Greek letters
$\alpha, \b, ...$. Given a walk $w_{2s}$, one can reconstruct the
paths from ${\cal C}(w_{2s})$ by assigning to the vertices of
$g(w_{2s})$ different values  from the set $\{1,2,\dots, n\}$.
This procedure will be considered in more details in Section 3.

Let us determine $\CE(g_{2s})$. The edge $e=(\a,\b)$ is present in
$\CE(g_{2s})$ if and only if there exists an instant
 $t$, $0< t\le  2s$ such that $w_{2s}(t-1) =\a$ and $w_{2s}(t) = \b$. In
this case we will write that $e(t'+1) = (\a,\b)$. If $t'$ is the
instant of the first arrival to $\b$, we say that the vertex $\b$
and the edge $(\a,\b)$ are  created at the instant $t'$. In
general, the graph $g(w_{2s})$ is a {\it multigraph} because the
couple $\a, \b$ can be connected by several edges of
$\CE(g_{2s})$; these could be  oriented as $(\a,\b)$ or $(\b,\a)$.
We will  denote by $\vert \a,\b \vert$ corresponding non-oriented
edges. We can easily pass to the graph $\hat g = (\CV,\hat \CE)$,
where the set of non-oriented edges $\hat \CE$ contains the
couples $\{a,\b\}$ such that $\a$ and $ \b$ are joined by elements
of $\CE$. In this case we denote the corresponding element of
$\hat \CE$ by $[\a,\b]$.

Obviously, $\vert \CE(g_{2s})\vert =2s$. We will say that the
number of non-oriented edges $\vert \a,\b\vert$ determines the
number of times that the walk $w_{2s}$ passes  the edge
$[\alpha,\b] $.

We denote by $m_w(\a,\b;t)$ the multiplicity of the non-oriented
edge $[\a,\b]$, or in other words, the  number of times that the
walk $w$ passes  the edge $[ \a,\b]$ up to the instant $t$, $1\le
t\le 2s$:
$$
  m_w(\a,\b;t) = \#\left\{ t'\in [1,t]: (w(t'-1),
    w(t')) = (\a,\b) {\hbox{\ or\ }}
  (w(t'-1),  w(t')) = (\b,\a)
  \right\}.
  $$
Certainly, this number  depends on the walk $w_{2s}$ but we will omit
the subscripts $w$.

\V

As we have seen from (2.1), the paths and the walks we consider are closed by definition,
that is $w_{2s}(2s) = w_{2s}(0)$. There is another important
 restriction for the paths and walks we consider.
It follows from the fact that the probability distribution of
$a_{ij}$ is symmetric: \v \noindent $\bullet $ the weight
$\CQ(I_{2s})$ is non-zero if and only if each edge from
 $ \hat \CE(w_{2s})$
 is passed by $w_{2s}$ an even number of times.
 \v

\noindent In this case we will say that the path $I_{2s}$ and the
corresponding walk $w(I_{2s})$ are {\it even}. We see that our
studies concern  the even closed paths and even closed walks only.
Then $g_{2s}$ is always a multigraph and the following equality
holds
$$
m_w(\a,\b;2s) = 0(\hbox{mod }2).
 \eqno (2.3)
$$

It should be noted that the requirement (2.3) concerns the case
when random matrices $A_n$ are real symmetric. In the case of
hermitian matrices corresponding condition is more restrictive and
requires that each edge is passed an even number of times in the
way that the numbers of  "there" and "back" steps are equal. We
will call such walks as the double-even walks. It is easy to see
that all definitions and statements of the present section do not
change when switching from the even to the double-even walks.
 In
the present paper, we do not consider the case of hermitian
matrices in details.

We denote by $\CW_{2s}$ the set of all possible minimal even
closed walks of $2s$ steps. The next subsections are devoted to
the further classification of its elements.

\subsection{Closed and non-closed instants of self-intersections}

Given $w_{2s}\in \CW_{2s}$, we say that the instant of time $t$
with $w(t)=\b$ is {\it marked} if the walk has passed  the edge $[
\a,\b] $ with $\a = w(t-1)$ an odd number of times during the time
interval $[0,t]$, $t\le 2s$;
$$
m_w(\a,\b;t)=1(\hbox{mod }2), \quad \a=w(t-1),\ \b = w(t).
$$


\begin{figure}[htbp]
\centerline{\includegraphics[width=7cm]{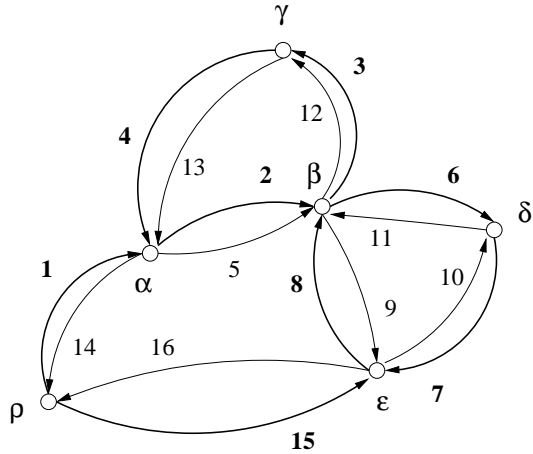}}
\caption{\footnotesize{ A graph  $g(w_{16})$ of the walk $w_{16}$
with two open and one closed instants of self-intersections}}
\end{figure}

In this case we will also say that the step $(t-1,t)$  and the
corresponding oriented edge $(\a,\b)=(w(t-1),w(t))\in \CE$ are
marked. Other instants of time are referred to as the {\it
non-marked} ones.

On Figure 1 we present an example of a minimal walk where the marked instants and
corresponding edges
are given in boldface.

Given a vertex $\b$ of the graph $g_{2s}$, we determine  the set
of all marked steps $(t_i,t_i+1)$ such that $w_{2s}(t_i)=\b$. The
set of corresponding edges  of the form
$(\b,\g_i)=(w_{2s}(t_i),w_{2s}(t_i+1))\in \CE$ is called the
 {\it exit cluster}
  of $\b$; we denote it by ${\cal D}_e(\b,w_{2s}) = {\cal
D}_e(\b)$. The cardinality $\vert {\cal D}_e(\b)\vert =
\deg_e(\b)$ is called the exit degree of $\b$. If $\vert
\CD_e(\b)\vert =0$, then we will say that the exit cluster of $\b$
is empty.

Each even closed walk $w_{2s}$ generates a binary sequence
$\theta_{2s} = \theta(w_{2s})$ of $2s$ elements $0$ and $1$ that
correspond to non-marked and marked instants, respectively. It is
clear that $\theta_{2s}$ represents path of $2s$ steps known as
the Dyck path \cite{Stan}. We denote by $\Theta_{2s}$ the set of
all Dyck paths of $2s$ steps.

Given $\b\in \CV(g_{2s})$, let us denote by $1\le t_1^{(\b)} < \dots < t^{(\b)}_N\le 2s-1$
the marked instants of time such that $w_{2s}(t^{(\b)}_j) = \b$.
We call $t^{(\b)}_j, 1\le j\le N$ the
{\it marked arrival instants} at $\b$.
 The {\it non-marked arrival instants} at $\b$ are defined in obvious manner.
 We will also say  that the step $(t_i^{(\b)}-1, t_i^{(\b)})$ and the
corresponding edge $e(t^{(\b)}_i)\in \CE$  are the {\it  arrival
step} at $\b$ and the {\it arrival edge} at $\b$, respectively. If
$N= 2$, then the corresponding vertex is called the vertex of
simple self-intersection \cite{SS1}. If $N=k$, then we say that
$\b$  is the vertex of $k$-fold self-intersection and that the
self-intersection degree of $\beta$ is equal to $k$; we denote the
self-intersection degree of $\b $ by
$\kappa(\b)=\kappa_{w_{2s}}(\b)$.

As it is mentioned in \cite{S}, one has to   consider the origin
of time $t=0$ as the marked instant of time. This is needed to
include the walks of the form $(1,2,3,1,2,3,1)$ with $\b=\{1\}$
and only one marked arrival $t_1^{(1)}=3$ into the family of walks
with self-intersections. Let us note that such "hidden" marked
instants of time can differ from $t=0$ and can be numerous. For
example, this happens each time when the walk returns at its
origin with all of the existing edges  closed. Summing up, we
accept that $\kappa(\a)\ge 1$ for any $\a\in \CV(g_{2s})$.

\vskip 0.3cm

The following definition generalizes the notion of the open vertex
of (simple) self-intersection introduced in \cite{SS2} and used in
\cite{R} and \cite{S}.

 \V
\noindent    {\bf Definition 2.1.}
{\it The instant $t$ is called the
\mbox{{non-closed (or open) arrival  instant}} at the vertex
 $\b \in \CV (g(w_{2s}))$, if  the step $(t-1,t)$ with $\beta = w_{2s}(t)$
 is marked and
if there exists at least one non-oriented edge $[ \b,\g]\in \hat
\CE$
 attached to $\b$ that is  passed an odd number of times
during the time interval $[0,t-1]$;
$$
m_w(\b,\g;t-1) = 1({\hbox{mod }} 2).
$$
In this case we say that the edge $[\b,\g]$ of the graph $\hat
g(w_{2s})$ is open up to the arrival instant $t= t^{(\b)}$, or
more briefly that this edge is $t$-open. The instant $t$ can be
also  called the open instant of self-intersection.
Correspondingly, one can define the $t$-open vertex $\b$ of
self-intersection of the walk $w_{2s}$. }

\V {\it Remarks}

1. Definition 2.1 remains valid in the case when $\g$ coincides with
 $\b=w_{2s}(t)$, more precisely in the case when
there exists another marked  instant $t'<t$ such that $\b = w_{2s}(t'-1) = w_{2s}(t')$
and the graph $g(w_{2s})$ has a loop
at the vertex $\beta$.  Definition 2.1 is also valid in the case when
 $\b = w_{2s}(0) = w_{2s}(t)$.

\v  2. Both of the definitions of the open vertex of self-intersection \cite{SS2} and the
open arrival instant are based on the following property:
the walk, when arrived at $\b$ at such  a marked instant,
has more than one possibility to continue its way with the non-marked step.
In the opposite case,
when  the only one  continuation
with the non-marked step is possible, we say that the arrival instant is closed.
A vertex of a walk can
change the property to be closed or open several times during the run of the walk.
On Figure 1 we present an example of the walk
 where two instants of self-intersection are open (these are $t=4$ and $t=8$)
and one instant of self-intersection $t=15$ is closed.

\V Regarding  the simple self-intersections only, we see that the
\mbox{definition 2.1} coincides with the definition of the open
vertex of simple self-intersection formulated first in
\cite{SS2,S} and then used in \cite{R}. However, the definition of
the open vertex of simple self-intersection presented in \cite{R}
slightly differs from that of \cite{SS2}. In \cite{R}, the vertex
$\b$ of the simple self-intersection at the marked instant $t$ is
called the open one when the edge $(\a,\b)$ of the first arrival
to $\b$ is not used again up to the instant $t$, while in
\cite{SS2} only returns in the direction $(\b,\a)$ are prohibited.
Looking at the Figure 1, we see that
 $\b=w_{16}(2)$ is the open vertex of the self-intersection
at the instant $t=8$ according to the definition of \cite{SS2},
and is not the open vertex according to the definition of
\cite{R}. However, as we will see later, this slight divergence in
definitions does not alter much the estimates one obtains (see
subsection 3.4).

\subsection {Primary and imported cells}

Given a walk $w_{2s}$, let us consider the following procedure of
reduction that we denote by $\cal P$: find an instant of time
$1\le t< 2s $ such that the step $(t-1,t)$ is marked and
$w_{2s}(t-1) = w_{2s}(t+1)$; if it exists, consider a new walk
$w'_{2s-2} = {\cal P}(w_{2s}) $ determined by a sequence
$$
w'_{2s} = (w_{2s}(0), w_{2s}(1), \dots , w_{2s}(t-1), w_{2s}(t+1), \dots, w_{2s}(2s)).
$$
Performing this procedure once more, we get $w''_{2s} = {\cal
P}(w_{2s}')$. We repeat  this procedure as many times as it is
possible and denote by $\bar W(w_{2s})$ the walk obtained as a
result when all of the reductions are performed.  Let us note that
$\bar W(w_{2s})$ is again a walk that can be transformed into the
minimal one by renumbering  the values of $\bar W(t)$, $t\ge 1$.
Then one can construct the graph
 $\bar g(w_{2s})= g(\bar W(w_{2s}))$ as it is done in subsection 2.1.

\noindent We accept the point of view when the graph
 $\bar g(w_{2s})$ is considered as a sub-graph of $g(w_{2s})$
$$
\CV(\bar g(w_{2s}))\subseteq \CV(g(w_{2s})),\quad
 \CE(\bar g( w_{2s}))\subseteq \CE(g(w_{2s}))
$$
and assume that the edges of $\bar g(w_{2s})$  are ordered
according to the order of the edges of $g(w_{2s})$.

\v {\bf Definition 2.2}. {\it Given a walk $w_{2s}$, we consider a
vertex of its graph $\b\in \CV(g(w_{2s}))$ and refer to the marked
arrival edges  $(\a,\b)\in \CE(g(w_{2s}))$ as to  the  primary
cells of $w_{2s}$ at $\b$. If $\b\in \CV(g(\bar W))$ with $\bar W
= \bar W(w_{2s})$, then we call the non-marked arrival edges
 $(a',\b)\in \CE(g(\bar W))$ the imported cells of $w_{2s}$ at $\b$. }

\v As we will see later, in order to control the exit degree of a
vertex of a walk with typical $\theta$, one needs to take into
account the number of imported cells at this vertex. The main
observation here is that the presence of the imported cells is
closely related with the breaks of the tree structure performed by
the walk. To formulate the rigorous statement, we need to
introduce the notion of the instant of broken tree structure that
will play the crucial role in our studies.

\v {\bf Definition 2.3.} {\it Any walk  $\bar W=\bar W(w_{2s})$
contains at least one instant $\bar \eta $ such that the  step
$(\bar \eta-1,\bar \eta)$ is marked and the step
 $(\bar \eta, \bar\eta+1)$ is not.
 We call such an instant $\bar  \eta$ the instant
of broken tree structure (or the BTS-instant of time) of the walk
$\bar W$. Passing back to the non-reduced walk $w_{2s}$, we
consider the edge $e(\eta)$ that corresponds to the edge
 $e(\bar\eta)\in \CE(\bar W)$ and refer to the instant $\eta$ as the
BTS-instant of the walk $w_{2s}$. }

\begin{figure}[htbp]
\centerline{\includegraphics[width=5cm]{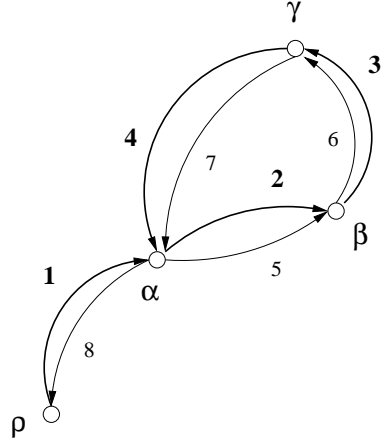}}
\caption{\footnotesize{ The graph  $g(\bar W)$ of the reduced walk
 $\bar W= \bar W(w_{16})$ }}
\end{figure}

{\it Examples}. Let us consider the  Figure 2, where we present
the graph of the walk
 $\bar W_8  = \bar W(w_{16})$ obtained from the walk $w_{16}$ given on Figure 1
 as a result of four reductions ${\cal P}$. Then
 each non-marked edge of the graph $g(\bar W)$ represents an
imported cell with respect to the vertices of $g(w_{2s})$. One can
make the following observations.

$\bullet$ The vertex $\b \in \CV(\bar g)$ has one non-marked edge
attached to it, so
 there is one imported cell at $\b$ of $g(w_{16})$; this is given
 by the edge $e(5)$. There are two primary cells at $\b\in \CV(g)$
 given by the edges
 $e(2)$ and $e(8)$.
 The root vertex $\varrho\in \CV(g)$ has one imported cell $e(14)$
and the vertex $\varepsilon\in \CV(g)$ has two primary cells,
$e(7)$ and $e(15)$, and no imported cells.

$\bullet$ The walk $\bar W_8$ has only one BTS-instant, $\bar
\eta=4$ because the edge $e(4)$ is marked and the edge $e(5)$ is
not.
 The same is true for the non-reduced walk
$w_{16}$ where $\eta=\bar \eta =4$. However, it can happen that in
the non-reduced walk $w_{2s}$ the marked BTS-instant $\eta$ is
separated from the corresponding non-marked instant by a tree-type
sub-walk that is to be removed during the reduction procedure. We
will not give the rigorous definition of the tree-type sub-walk
because we do not use it here in the full extent.

$\bullet$ Returning to the graph of $w_{16}$ depicted on Figure 1,
we can explain the term "imported cell" on the example of the
vertex $\b=w_{16}(2)$. The exit cluster of this vertex consists of
two edges, $e(3)$ and $e(6)$. Regarding the tree $\CT(w_{16})$, we
see that these  edges have different parents in $\CT(w_{16})$; the
edge of the tree that corresponds to $e(3)$ has the origin
determined by the end of the edge corresponding to $e(2)$ while
the edge $e(6)$ is imported at $\b$ by the edge $e(5)$ from
another part of the tree $\CT(w_{16})$. We give more examples of
the walks with primary and imported cells at the end of this
paper.

 \v  Now we will present a  proposition that trivially
follows from the Definitions 2.1 and 2.3. However, it plays so
important role in our studies that we formulate it as the separate
statement.

\v {\bf Lemma 2.1.} {\it  If the arrival instant $ \t$ is the
BTS-instant of the walk $w_{2s}$, then $\t$ is the open instant of
self-intersection of $w_{2s}$.}

\V Given a vertex $\b\in \CV(w_{2s})$, we refer to the
BTS-instants $\eta_i$ such that $w_{2s}(\eta_i)= \b$ as to the
$\b$-local BTS-instants. All other BTS-instants are referred to as
the $\b$-remote BTS-instants.
 Now we can formulate the main result of this
section.

\v {\bf Lemma 2.2.}
 {\it Let $ K^{(\setminus \b)}_{w_{2s}}$ be the number
of all $\b$-remote  BTS-instants \mbox{of the walk $w_{2s}$}. Then
the number of all imported cells at $\b$  denoted by
$J_{w_{2s}}(\b)$ is bounded as follows}
$$
J_{w_{2s}}(\b) \le  K^{(\setminus \b)}_{w_{2s }} +
\kappa_{w_{2s}}(\b).
 \eqno (2.4)
$$

\v {\it Proof.} Let us consider the reduced walk
 $\bar W = \bar W(w_{2s})$ and a vertex $\b\in \CV(g(\bar W))$. We introduce the
function $\Lambda_\b(t;\bar W)$ determined as the number of
$t$-open edges attached to $\b$;
$$
\Lambda_\b(t;\bar W) = \# \{i: m(\a_i,\b;t) = 1 ({\hbox{mod }} 2)
\}
$$
and consider how this function changes its values at the instants
of time when $\bar W$ arrives at $\b$. The following
considerations show that this value can be changed by $ 0,+2$ and
$ -2$ only.

I. The case when the value of $\Lambda_\b$  stays unchanged is
possible in two situations.

a) The first situation happens when  the walk $\hat W$ leaves $\b$
by a non-marked edge and arrives at $\b$ by a marked edge. Then
the corresponding cell is the primary one and we do not care about
it.

b) The second situation occurs when the walk $\hat W$ leaves $\b$
by a marked step $(x,x+1)$ and arrives at $\b$ by a non-marked
step $(y-1,y)$. Then the interval of time $[x+1,y-1]$ contains at
least one BTS-instant of time. This is the first instant $\tau$
when the non-marked step follows immediately after the marked one.
It is clear that $w_{2s}(\tau)\neq \b$ and therefore $\tau$ is the
$\b$-remote BTS-instant.

It should be noted that another such interval $[x'+1,y'-1]$
contains another $\eta'$ that obviously differs from $\eta$;
$\eta'\neq\eta$.  This is because the any couple of such time
intervals $[x+1,y-1]$, $[x'+1,y'-1]$ has an empty intersection.
Then each imported cell of the type (Ib) has at least one
corresponding $\b$-remote BTS-instant and the sets of the
BTS-instants that correspond to different intervals do not
intersect.

II. Let us consider the arrival instants at $\b$ when the value of
$\Lambda_\b$ is changed.

a) The change by $+2$ takes place when the walk $\bar W$ leaves
$\b$ by a marked edge and arrives at $\b$ by a marked edge also.

b) The change by $-2$ occurs in the opposite case when $\hat W$
leaves $\b$ with the help of the non-marked edge and arrives at
$\b$ by a non-marked edge.

During the whole walk, these two different passages occur the same
number of times. This is because $\Lambda_\b=0$ at the end of the
even closed walk $\bar W$. Taking into account that the number of
changes by $+2$ is bounded by the self-intersection degree
$\kappa_{\bar W}(\b)$, we conclude that the number of imported
cells of this kind  is not greater than
 $\kappa_{\bar W}(\b)$.

To complete the proof, we have to  pass back from the reduced walk
$\bar W(w_{2s})$ to the original $w_{2s}$. Since the number of
imported cells of $w_{2s}$ and the number of BTS-instants of
$w_{2s}$ are uniquely determined by $\bar W (w_{2s})$, and
$$
\kappa_{\bar W}(\b)\le \kappa_{w_ {2s}}(\b),
$$
then (2.4) follows provided $\b$ belongs to $g(\bar W)$ as well as
to $g(w_{2s})$. If $\b\notin \CV(g(\bar W))$, then
$J_{w_{2s}}(\b)=0$ and (2.4) obviously holds.
 Lemma 2.2 is proved.
  $\diamond $

\v {\bf Corollary of Lemma 2.2.} {\it Given a vertex $\b$ of the
graph of $w_{2s}$, the number of primary and imported cells
$L(\b)$ at $\b$ is bounded
$$
L(\b)\le 2\kappa(\b) +K,
 \eqno (2.5)
$$
 where $K$ is the
total number of the BTS-instants performed by the walk $w_{2s}$.}

\v {\it Proof.} The number of the primary cells at $\b$ is given
by the $\kappa(\b)$. The number of the imported cells $J(\b)$ is
bounded by the sum $\kappa(\b) + K^{{(\setminus \b)}}$, where
$K^{(\setminus \b)}\le K$. Then (2.5) follows.
 $\diamond$

\v Let us note that relations (2.4) and (2.5) reflect that
non-local character of the imported cells that can arise at $\b$
due to the BTS-instants that can happen rather "far" from $\b$ at
any remote part of the walk. In Section 4 we show that these cells
can "bring" to $\b$ the edges that originally do not belong to the
corresponding  vertex in the initial tree $\CT_s$. This explains
the use of the term imported cells for the corresponding edges (or
instants of time).

The results we have formulated and proved in this section are
sufficient to obtain the estimates we need for the high moments of
large random matrices (see Section 4). So, at this stage we
terminate the study of the fairly rich and interesting subject
given by the family of even closed walks.



\section{Estimates of the set of paths}

In the previous section, we have described the properties of the
even closed paths $I_{2s}$ based on the notion of the vertices and
the instants of self-intersections. In the present section we
describe the procedure of construction of the set of paths and
present corresponding estimates. Here we mostly follow the scheme
of \cite{SS1} and \cite{SS2} added by considerations of the
vertices of open self-intersections that are not necessary simple.

\subsection{Classes of equivalence}

To describe  the constructions procedure, we first complete and
summarize the description of the even walk paths $\CI_{2s}$
started in the Section 2. There a way to separate
 the set of all paths $\CI_{2s}(n)$
 into classes of equivalence $\CC (w_{2s})$ was considered.

Each class of equivalence
 ${\cal C}(w_{2s})$ is uniquely determined by an element $w_{2s}\in
\CW_{2s}$. It is clear  that the weights of the equivalent paths
are also equal; if  $I_{2s}\sim I'_{2s}$, then
 ${ \CQ} (I_{2s}) = { \CQ} (I'_{2s})=  \CQ(w_{2s})$.
Therefore we can rewrite (2.1) in the form
$$
\E \left\{\T A_n^{2s}\right\} = \sum_{w_{2s}\in \CW_{2s}} \vert
{\cal C}(w_{2s}) \vert \cdot
 { \CQ}(w_{2s})
  \eqno (3.1)
$$
Clearly,  the cardinality of the class ${\cal C}(w_{2s})$ is
determined by the number of vertices of the graph $g(w_{2s})$; $
\vert {\cal C}(w_{2s}) \vert = n(n-1)\cdots (n-\vert
\CV(g_{2s})\vert +1)$.

Regarding $g(w_{2s})$, we determine the partition of $\CV(g)$ into
subsets $\CN_2$, \dots, $\CN_s$; if  $\kappa(\a)=k$, then $\a\in
\CN_k$.  We denote by $\nu_k$ the cardinality of $\CN_k$,
$$
\nu_k = \vert \CN_k\vert, \quad \nu_k\ge 2.
$$
Denoting by $\CN_1$ the subset of of vertices $\a$ of $g(w_{2s})$
with $\kappa(a)=1$,  we get obvious equality $s = \sum_{k=1}^s
k\nu_k$, where $\nu_1=\vert \CN_1\vert$.

Given a walk $w_{2s}$ that has $\nu_k$ vertices of $k$-fold
self-intersections, $2\le k\le s$, we  say that it is of the type
$\bar \nu_s= \bar \nu(w_{2s}) =  (\nu_1,\nu_2,\dots,\nu_s)$.
 Then  we can separate the set of all walks $\CW_{2s}$  into classes of
 equivalence.
We say that two  walks $w_{2s}$ and $w'_{2s}$ are equivalent,
$w_{2s}\sim w'_{2s}$ if their types of self-intersections
coincide, $\bar \nu (w_{2s}) = \bar \nu (w_{2s}')$. The number
$$
\vert \bar \nu_s\vert_1 = \sum_{k=2}^s (k-1)\nu_k
 \eqno (3.2)
$$
determines the cardinality of $\CV(w_{2s})$; namely,
 $\vert \CV(w_{2s}) \vert = s +1 - \vert \bar \nu_s\vert_1$.
We consider the  rearrangement of (3.1) according to the classes
of equivalence  described above in the next subsection.

Finally, let us recall that the walk $w_{2s}$ generates a Dyck
path $\theta_{2s} = \theta(w_{2s})$ that is in one-by-one
correspondence with a sequence of $s$ marked instants $\Xi_s =
(\xi_1,\dots,\xi_s)$ such that $\u_1<\u_2<\dots<\u_s$ and $\u_j\in
\{1,2s-1\}$. As before, here it is convenient to consider the
subscript of $\u_j$ as a sort of the discrete "time" we denote by
$\t$, $1\le \t\le s$. Sometimes we will refer to $\tau$ as to the
instants of $\tau$-time, or more simple as to the $\tau$-instants.
Given a value of $\tau$, we say that $\u_\tau = \U_s(\tau)$.

Remembering the definition of the vertex $\a$ of $k$-fold
self-intersection of the walk $w_{2s}$, we see that it is
determined by an ordered $k$-plet of variables
$\tau^{(1)}<\dots<\tau^{(k)}$ such that the marked instants
$\u_{\tau^{(1)}}<\dots<\u_{\tau^{(k)}}$ indicate the marked
arrival instants at the vertex $\a$ determined by the first
arrival  $\a = w_{2s}(\u_{\tau^{(1)}})$. We also observe that the
elements $\a_j\in \CN_k$, $1\le j\le \nu_k$ are naturally ordered
in the chronological  way. Therefore the walk $w_{2s}$ generates
an ordered set of ordered $k$-plets that we denote by $
T^{(k)}(w_{2s}) $.

More generally, we denote the ordered set of $\nu_k$ ordered
$k$-plets by
$$
 T_s^{(k)}(\nu_k)=\{
(\tau_j^{(1)}(k),\dots,\tau_j^{(k)}(k)),\,
 1\le j\le\nu_k\},
 $$
where all values $\tau_j^{(l)}(k)$ are distinct. In what follows,
we omit variable $k$ in $\tau_j^{(l)}(k)$ when regarding such
$k$-plets. Also, we will use denotations $(\tau',\tau'')$ for the
$\tau$-instants of simple self-intersection.

The last observation concerns the fact that each instant of
self-intersection can by characterized by one more property - is
the corresponding arrival instant open or not. So, the variable
$\tau^{(l)}_j,l\ge 2$ should be assigned by one more subscript
that in the binary way indicates the openness of
$\u_{\tau^{(l)}_j}$. An obvious but important remark is that the
openness of $\u_{\tau^{(l)}_j}$ depends on the whole pre-history
of walk $w_{2s}$; that is on its sub-walk determined by the time
interval $[1,\u_{\tau^{(l)}_j}-1]$.



\subsection{Cover the set of walks}

Let us recall that the walk $w_{2s}$ of $2s$ steps is determined
as a sequence of $2s+1$ symbols (labels or numbers). To determine
a particular walk, one starts with the initial (root) label and
then indicates  the symbols that appear at the subsequent instants
of time $t$, $1\le t\le 2s$.

In the case of  closed even walk one needs less information. The
first condition implies relation $w_{2s}(2s)=w_{2s}(0)$ and the
second says that, for instance, if $w_{2s}$ has no
self-intersections, then it is sufficient to indicate the values
of $w_{2s}(t)$ at $s$ instants of time. Indeed, elementary
reasoning shows that this choice of values corresponds to the
choice of one of the Dyck paths from the set $\Theta_{2s}$. Then
the choice of labels at $s$ instants of time corresponds to the
choice of the marked instants. The values of $w_{2s}(t)$ at each
non-marked instant of time are uniquely determined by the rule
that the non-marked step $(\b,\a)$ closes the last marked arrival
$(\a,\b)$.  This arrival  is unique in the case when $w_{2s}$ has
no self-intersections.

The  rule described above uniquely determines the even walk by the
knowledge of the Dyck path $\theta_{2s}$ and the partition
 $\bar \CN_s=  (\CN_1,\CN_2,\dots,\CN_s)$. We call this rule the
canonical run and refer to such a walk as to the tree-like walk.

The situation becomes more complicated in the case when we want to
determine the set of all even  walks with a number of
self-intersections.
 The simplest example is given by
the following two walks of $2s=8$ steps
$$
\tilde{w}_8 = (1,2,3,4,2,4,3,2,1)\quad {\hbox{and}}\quad
\breve{w}_8= (1,2,3,4,2,3,4,2,1)
$$
that have the same Dyck path $\theta_8= (1,1,1,1,0,0,0,0)$ and the
same partitions with $ \CN_1 = \{3,4\}$ and  $\CN_2 =\{2\}$. The
walk $\tilde w_8$ is the tree-like walk but $\breve w_8$ is not of
the tree-like structure.

We denote the set of minimal even walks that have the same
$\theta_{2s}$ and $\bar \CN_s$ by $\CW(\theta_{2s},\bar \CN_s)$
and denote by $\bar \CN_s^{(r)}$ with $r$, $0\le r\le \nu_2$ open
vertices of simple self-intersections.

 It is argued in
\cite{SS2} that given $\theta_{2s}$, the cardinality of
$\CW(\theta_{2s},\bar \CN_s)$ is bounded as follows;
$$
\vert \CW(\theta_{2s},\bar \CN_s^{(r)})\vert \le W(\bar\nu_s,r) =
3^{r} \prod_{k=3}^{s} (2k)^{k\nu_k}.
 \eqno (3.3)
$$
 Let us prove that (3.3) is true. To do this, we
will need the following simple statement.

\v
 {\bf Lemma 3.1.}
  {\it Consider a vertex $\b$ with the self-intersection degree $\kappa(\b) =
k$. The total number of non-marked edges of the form $(\b,\a_i)$
is equal to $k$ and   at any instant of time $1\le t\le 2s$, the
number of $t$-open marked edges attached to $\b$ is bounded by
$2k$. }

\v {\it Proof.} To prove the first part of this lemma,  we denote
by $\I_m$ and $\I_{nm}$ the numbers of marked and non-marked
enters at $\b$ and by $\O_m$ and $\O_{nm}$ the numbers of marked
and non-marked exits from $\b$, respectively.

The walk is closed and therefore number of enters at $\b$ is equal
to the number of exits from $\b$. Since the walk is even, then the
number of the marked edges attached to $\b$ is equal to the number
of non-marked edges attached to $\b$. Corresponding equalities,
$$
\I_m + \I_{nm} =\O_m + \O_{nm}, \quad {\hbox{and}}\quad \I_m +\O_m
= \I_{nm} + \O_{nm}
$$
result in $2(\I_m - \O_{nm})=0$. Then $\O_{nm}=k$.

Given $t$, we consider the numbers $\I_m(t), \I_{nm}(t), \O_m(t)$,
and $\O_{nm}(t)$ that count the corresponding numbers of edges
during the time interval $[0,t-1]$. If $t$ is the instant of the
exit from $\b$, then
$$
\I_m(t) + I_{nm}(t) =\O_m(t) + \O_{nm}(t) + \varphi,
\eqno
(3.4)
$$
where $\varphi = 0$ in the cases when $\b$ is the root vertex and
$\varphi=1$ otherwise. The choice of edges to close by the
non-marked exit edge is bounded by the number of marked edges that
enter and leave $\b$, that is by the number $\I_m + \O_m - \I_{nm}
- \O_{nm}$. Taking into account (3.4), we get inequality
$$
\I_m + \O_m - \I_{nm} - \O_{nm}\le 2\left(
\I_m-\O_{nm}\right)-\varphi \le 2\I_m-\varphi\le 2k.
$$
Lemma 3.1 is proved. $\diamond$

\V

Lemma 3.1 says that the multiplicative contribution of each vertex
$\b$ to the estimate (3.3) is bounded by
$(2\kappa(\b)-\varphi)!!$. Let us recall that in the case when the
root vertex $\varrho$ has $l$ marked edges $(\g_i,\rho)$ attached,
then we accept that $\kappa(\varrho)=l+1$ because of the first
marked arrival instant at $\varrho$ that is hidden.

In the case of $\kappa(\b)=2$ the bound $4-\varphi$ can be
immediately improved. If the second arrival instant at $\b$ is
closed, then there is only one possibility to continue the walk at
the non-marked instant. If the second arrival instant in open,
then we have not more than three possibility to continue the run
at the non-marked departure from $\b$ in the case when
$\b\neq\varrho$ and not more than two possibilities in the case
when $\b=\varrho$. This terminates the proof of (3.3).

In fact, the estimate (3.3) can be further improved by the
analysis of the number of BTS-instants at the vertices of high
self-intersection degree. This goes out of the frameworks of the
present paper.

\subsection{The choice of instants of self-intersections}

In this subsection we estimate the number of possibilities to
choose $\nu_k$ instants of $k$-fold self-intersections that is
obviously bounded by the number of possibilities  to choose the
set of $k$-plets $T_s^{(k)}(\nu_k)$.
 We denote the latter number by  $\vert
T_s^{(k)}(\nu_k)\vert$. In this subsection  we do not do the
distinction between the open and the closed arrival instants.

\V {\bf Lemma 3.2.}
 {\it Given any $\theta_{2s}$,
the number of possibilities
  the number of possibilities to choose $\nu_2$
instants of simple self-intersections  is bounded by
$$
\vert T^{(2)}_{s} (\nu_2 ) \vert \le {1\over \nu_2!}\cdot
 \left({s^2\over 2}\right)^{\nu_2}.
 \eqno (3.5)
$$

If $k\ge 3$, then the number of possibilities
  the number of possibilities to choose $\nu_k$
instants of $k$-fold self-intersections  is bounded by
$$
\vert T^{(k)}_{s} (\nu_k ) \vert \le {1\over \nu_k!}\cdot
 \left({s^k\over (k-1)!}\right)^{\nu_k}.
 \eqno (3.6)
$$
 }

\v

{\it Proof.} Let us start with the proof of (3.5). Here we mostly
repeat the computations of \cite{SS2}. Regarding  the second
arrival instant at the vertex $\b_j$ of simple self-intersection,
we have to point out its position $\tau^{(2)}_j$ among the  $s$
marked instants, $1\le j\le \nu_2$. Denoting for simplicity
$\tau^{(2)}_j = l_j$, we can write that
$$
\vert T_s^{(2)}(\nu_2)\vert \le \sum_{1\le l_1< l_2<\dots<
l_{\nu_2} \le s} (l_1-1)_+ \ (l_2-3)_+\ (l_3-5)_+\ \cdots
(l_{\nu_2} - 2{\nu_2} +1)_+,
 \eqno(3.7)
$$
where  $(x)_+$ is equal to $x$ if $x\ge 0$ and to zero otherwise.
Indeed, the first factor of (3.7) reflects the fact that at the
$\tau$-instant $\tau_1^{(2)} = l_1$ we dispose of $l_1-1$ marked
instants of $\U(\theta_{2s})$ to choose the vertex $\b_1$ of this
self-intersection.  In other words, we have to choose the first
$\tau$-instant $\tau^{(1)}_1$ that creates
 $\b_1 =w_{2s}(\xi_{\tau_1^{(1)}})$. Regarding the second arrival
$\tau$-instant of the next simple self-intersection $\tau^{(2)}_2
= l_2 $, we can choose among $l_2-1-2$ marked instants to point
out the vertex of the second self-intersection. Here by
subtraction of $2$, we avoid the instants of the first
self-intersection already used. We proceed like this up to
$\nu_2$-th self-intersection.

Changing variables $l_i-1 = \hat l_i$, we derive from (3.7) that
$$
\vert T_s^{(2)}(\nu_2)\vert \le \sum_{1\le \hat  l_1<  \dots <
 \hat l_{\nu_2} \le s-1 }  \hat l_1  \hat  l_2 \cdots  \hat  l_{\nu_2} \le
 {1\over \nu_2 !} \sum_{1\le   \hat l_1,  \dots ,  \hat  l_{\nu_2}\le s-1}
  \hat  l_1  \hat l_2\cdots   \hat l_{\nu_2}\le
 {1\over \nu_2 !} \left(\sum_{ \hat l_1=1}^{s-1}  \hat  l_1\right)^{\nu_2}.
$$
Then  (3.5)  follows.

To prove (3.6), let us consider first the case of $\nu_k=1$. To
determine a vertex $\a$ of $k$-fold self-intersection, we have to
point out $k$ marked arrival instants at this vertex. Denoting
corresponding $\tau$-instants by $\tau^{(i)}$,
 $1\le i\le k$
  we see that
$$
 \vert T^{(k)}_s(1) \vert \le \sum_{1\le \tau^{(1)}< \dots<
\tau^{(k)}\le s} \ 1.
$$
This relation gives an obvious estimate
$$
\vert T^{(k)}_s(1)\vert \le {s \choose k} = {s (s-1)\cdots
(s-k+1)\over k!}\le {s^k\over k!}.
$$

To study the case of $\nu_k\ge 2$ vertices $\a_j$, $1\le j\le
\nu_k$ of $k$-fold self-intersections with $k\ge 3$, we use the
estimate of $T^{(k)}_s(1)$ with $k$  replaced by $k-1$. Namely,
pointing out the last arrival $\t$-instant $\tau^{(k)}$, we can
write that
$$
\vert T^{(k)}_s(1)\vert = \sum_{\tau^{(k)}=k}^s\vert
T^{(k-1)}_{\t^{(k)}-1}(1)\vert \le \sum_{\t^{(k)}=k}^s
{\t^{(k)}-1\choose k-1}$$
$$
 \le \sum_{\t^{(k)}=k}^s {(\t^{(k)}-1)\cdots (\t^{(k)}-k+1)\over
 (k-1)!}
  \le {s^{k-1}\over (k-1)!}\, \sum_{\t^{(k)}=k}^s 1\
  \le {s^k\over (k-1)!}.
   \eqno (3.8)
$$
This estimate is slightly worse than (3.7) but it simplifies the
study of the  general case of $\nu_k\ge 1$.

Let us denote the last arrival instants at $\nu_k$ vertices
$\tau_{1}^{(k)}<\dots <\tau_{\nu_k}^{(k)}$. Using the
representation described in (3.8) and  ignoring the fact that some
of the marked instants are already in use, we get inequality
$$
\vert T^{(k)}_{s} (\nu_k )\vert \le \sum_{1<\tau_{k}^{(1)}<\dots
<\tau_{\nu_k}^{(k)}<s} \vert T^{(k-1)}_{\t^{(k)}_1-1}(1)\vert
\cdots \vert T^{(k-1)}_{\t^{(k)}_{\nu_k}-1}(1)\vert
$$
$$
 \le \left( {s^{k-1}\over (k-1)!}\right)^{\nu_k}\cdot
\sum_{1<\tau_{k}^{(1)}<\dots <\tau_{\nu_k}^{(k)}<s} 1\
 \le {1\over \nu_k!}\left( {s^{k}\over (k-1)!}\right)^{\nu_k}
 \eqno (3.9)
 $$
that gives (3.6). Lemma 3.2 is proved. $\diamond$

\V Let us note that our estimate of $T^{(k)}_s(1)$  and that of
(3.9) slightly differ from the estimates of $T^{(k)}_s(\nu_k)$
presented in \cite{SS2,S} and used in \cite{R}. In these works the
vertex $\a$ of $k$-fold self-intersection has been considered as
the vertex, where the first two arrivals at $\a$ produce a simple
self-intersection and then the remaining $k-2$ arrivals increase
the self-intersection degree $\kappa(\a)$ up to $k$. Our
representation (3.8) can be viewed as the backward procedure of
the choice that starts from the last arrival instant $\tau^{(k)}$
at $\a$ and ends with the first one $\tau^{(1)}$. Evidently, these
two descriptions coincide in the case of the simple
self-intersections (3.7). Our approach is in more agreement with
the scheme of (3.7). Also the backward scheme of the estimates
seems to be the most convenient  in the studies of vertices with
one or more open instants of self-intersections that we carry out
in the next subsections.

\subsection{Open and closed simple self-intersections}

Lemma 3.2 gives the following estimate for the total number of
possibilities
$$
\prod_{k=2}^s \vert T_s^{(k)}( \nu_k)\vert  \le {1\over \nu_2!}
\left({s^2\over 2}\right)^{\nu_2} \cdot  \prod_{k=3}^{s}\, {1\over
\nu_k!}
 \left({s^k\over (k-1)!}\right)^{\nu_k}.
\eqno (3.10)
$$
However, this estimate  cannot be used directly in this form. The
reason is in the presence of the factor $3^{r}$ in the estimate
(3.3) that increases too much the product of the right-hand sides
of these estimates. This obstacle was treated first in the paper
\cite{SS1}, where it was shown that the choice of the vertex of
open self-intersection is much less than that of the vertex of
closed simple self-intersection given by  $s^2/2$ as $s\to\infty$.
In the present subsection we describe the procedure of
construction of the open vertices of simple self-intersections by
formalizing  the arguments of \cite{SS1} and slightly modifying
them to take into account two different types of open
self-intersections.

Let us  recall that given $\theta_{2s}$, we can  construct a walk
with vertex of (simple) self-intersection by pointing out
 a couple of marked instants $(\xi',\xi'')$ from the set $\Xi(\theta_{2s})$
 or equivalently by pointing out the couple of $\tau$-instants
 $(\tau',\tau'')$. This means that we choose first $\tau''$ and
 then we point out the vertex of the self-intersection
 $\a=w_{2s}(\xi_{\tau'})$ by choosing $\tau'<\tau''$.

 Aiming the construction of the open vertex of simple self-intersection,
 we want to be sure that  $\a$ is such that there exists an edge $e'$ attached
 to $\a$ that is $(\xi''-1)$-open. The vertex $\a$ can be the end (or the
 head)
 of $e'$ and then we have the open self-intersection of the first type we call
 the E-open self-intersection; if $\a$
 represents the start (or the tail) of $e'$, then the self-intersections is of
 the second type, or briefly of the S-type.

 An important remark is to be made here. As we have seen, the
 openness of the edge depends on the particular run of the walk.
For example, regarding the Dyck path $\theta_{12}=
(1,1,1,1,0,0,1,1,0,0,0,0)$ with two self-intersections determined
by the couples of  instants $t$ of time $(1,4)$ and $(2,8)$, we
see that the instant $t''_2=8$ is the open or the closed instant
of self-intersection in dependence on the value of $w_{12}$ at the
non-marked instant $t=5$.

However, the number of choices of the edge $e'$ does not depend on
the particular run of the walk and is bounded by the number of
$(t''_2-1)$-open edges. This number is not greater than the value
$\theta_{2s}(t'')$.

\V Taking into account the previous reasoning, we describe the
following procedure of construction of the set of  walks with
closed and open simple self-intersections.

1. Given $\theta_{2s}$, we choose among $s$  $\tau$-instants the
positions of $\nu_2$ second arrivals
 $\tau''_1<\dots<\tau''_{\nu_2}$ at vertices of simple self-intersections. Among
them, we choose $\nu_2- r_E - r_S$ instants that will be the
instants of closed self-intersections, and $r_E$ and $r_S$
instants of E-open and S-open self-intersections, respectively. We
denote $r=r_E+r_S$.  With this information in hands, we start the
run of the walk.

 \v 2. At the instant $\tau''_1$ we have choose a
vertex of the first self-intersection. Then the following
estimates are valid:

\begin{itemize}
\item if
$(\tau'_1,\tau''_1)$ is prescribed to be the closed
self-intersection, then there is not more that $\tau''_1-1$
possibilities to choose $\tau'_1$;

\item if $(\tau'_1,\tau''_1)$ is prescribed to be the E-open
self-intersection, we choose the vertex among the heads of the
edges that are $(\xi_{\tau''_1}-1)$-open; the number of such edges
is not greater than $\theta_{2s}(\xi_{\tau''_1})$;

\item if $(\tau'_1,\tau''_1)$ is prescribed to be the S-open
self-intersection, we choose the vertex among the tails of the
$(\xi_{\tau''_1}-1)$-open edges; again, there is not more than
$\theta_{2s}(\xi_{\tau''_1})$ of such edges.
\end{itemize}

3. To continue the run if the walk at the instant
$\xi_{\tau''_1}+1$, we look whether it is marked or not. If it is
marked and is not the instant of self-intersection, we produce a
new vertex; if it is the instant of self-intersection, we return
to the paragraph 2 given above.

4. If the instant $\xi_{\tau''_1}+1$ is non-marked, and the
self-intersection $(\tau'_1,\tau''_1)$ is the closed one, then
vertex $w_{2s}(\xi_{\tau''_1}+1)$ is uniquely determined. If
$(\tau'_1,\tau''_1)$ is open, then we consider all possibilities
to choose the vertex $w_{2s}(\xi_{\tau''_1}+1)$; this produces
several sub-walks that correspond to different runs. In the
previous subsection we have seen that there are not more than
three possible runs in the case of simple open self-intersection.
Then we proceed with the next step $\xi_{\tau''_1}+2$ till we meet
the second arrival instant $\xi_{\tau''_2}$ of the second
self-intersection. At this stage we redirect  ourselves to the
paragraph 2.

We continue the procedure described until the last vertex of the
self-intersection $(\tau'_{\nu_2}, \tau''_{\nu_2})$ is determined.
In this construction, we take into account that the choice of
vertices of $j$-th closed self-intersection is bounded by
$l_j-2j-1$, where $l_j$ is the position of $\tau''_j$ in
$\Xi(\theta_{2s})$.

\V Having all the steps of the construction performed, we obtain
not more than $3^{r_E+r_S}$ different runs, and whatever run
 is chosen, there is not more than
$$
(l_1-1)\cdots (l_{\nu_2}-2\nu_2 -1) \cdot M_\theta^r, \quad
M_\theta = \max_t \theta_{2s}(t)
$$
possibilities to choose the instants and  vertices of the
self-intersections. We see that in the case of walks with simple
self-intersections, we have not more than
$$
\sum_{1\le l_1<\dots< l_s\le s} \ \ \sum_{(\nu_2-r,r_E,r_S)
\subset (1,2,\dots,s)} \ (l_{i_1}-1)\cdots
(l_{i_{\nu_2-r}}-2(\nu_2-r) -1) \cdot M_\theta^r,
$$
possibilities to choose the vertices of simple self-intersection
of the types indicated, where the second sum corresponds to the
choice of $\nu_2-r, r_E$ and $r_S$ vertices mentioned above. This
gives the estimate
$$
{1\over \nu_2!}\cdot  {\nu_2!\over (\nu_2-r)! \, r_E!\, r_S!} \ \
\left({s^2\over 2}\right)^{\nu_2-r} \cdot (sM_\theta)^r.
$$
Summing over all possible  values of  $r_E$ and $r_S$, we arrive
at the proof of the following statement.

 \v
{\bf Lemma 3.3}. {\it Given $\theta_{2s}$, the number of
possibilities $\vert T^{(2)}_s(\nu_2;r)\vert$ to construct $\nu_2$
vertices of simple self-intersections with $r$ open ones is
bounded as follows;
$$
\vert T^{(2)}_s(\nu_2;r)\vert\le {1\over (\nu_2-r)! \ r! } \ \cdot
(2sM_\theta)^r \left({s^2\over 2}\right)^{\nu_2-r} ,\quad 0\le
r\le \nu_2.
  \eqno (3.11)
$$
}

In Section 4, we use (3.10) modified with the help of (3.11).


\subsection{Triple self-intersections with open arrival instants}

Repeating the main lines of the previous subsection, it is easy to
estimate the number of possibilities to create a walk with  one
vertex of triple self-intersection that has  one or two open
arrival instants. Let us denote the arrival instants at this
vertex by $\tau^{(1)}< \t^{(2)}<\t^{(3)}$. We set up the value of
$\t^{(3)}$ and consider the following three possible scenarios
that can happen.
 \v
 {\bf i)} The first situation is given by the case
when the second arrival instant $\t^{(2)}$ is open and $\t^{(3)}$
is closed. In this case we can choose the position for $\t^{(2)}$
among $s$ marked instants of $\U(\theta_{2s})$ with the only
restriction that $\t^{(2)}<\t^{(3)}$.

We start the run of the walk $w_{2s}$ and at the instant
$\t^{(2)}$ we indicate an edge $e$ that already exists in the
graph of the sub-walk and that is $(\xi_{\t^{(2)}}-1)$-open. This
edge can be chosen from the set of cardinality less or equal to
$\theta_{2s}(\xi_{\t^{(2)}})$ that is bounded by $M_\theta$. We
set $w_{2s}(\xi_{\t^{(2)}})=\b$ to be equal to the head or to the
tail of $e$ and continue the run of the walk remembering that at
the $\tau$-instant $\t^{(3)}$ the walk has to visit the vertex
$\b$ already pointed out; $w_{2s}(\xi_{\t^{(3)}}) = \b$. The walk
is constructed and the number of possibilities to choose the
$\t$-instants $\tau^{(1)}$ and $ \t^{(2)}$ is bounded by
 $\t^{(3)} M_\theta$.

 \v
  {\bf ii)} Let us consider the case when the arrival instant
$\t^{(2)}$ is closed and $\t^{(3)}$ is open. Remembering that the
value of $\t^{(3)}$ is already set, we start the run of the walk.
At the instant $\t^{(3)}$, we choose the vertex $\b$ that is the
head or the tail of an edge that is $(\xi_{t^{(3)}}-1)$-open and
set $w_{2s}(\xi_{\t^{(3)}})=\b$. Then the $\t$-instant $\t^{(2)}$
is determined in the way that  $\b = w_{2s}(\xi_{\t^{(2)}})$ and
the choice of $\t^{(2)}$ is bounded by $M_\theta$. It remains to
attribute to $\t^{(1)}$ any value such that $\t^{(1)}<\t^{(2)}$
and we are done. Again we see that the number of possibilities to
create a couple $\t^{(1)}$ and $\t^{(2)}$ is bounded by
 $\t^{(3)} M_\theta$.

 \v
  {\bf iii)} Finally, let us construct a walk with triple self-intersection
   such that the both of the
instants $\t^{(2)}$ and $\t^{(3)}$ are the open ones. We still
assume that $\t^{(3)}$ is already determined. We start the run of
the walk and the question now is to choose two vertices $\b_1$ and
$\b_2$ that already exist at the instant $\xi(\t^{(3)})-1$. There
are two different cases.

\v

Let us  assume that $\b_2$ is such that the
$(\xi_{\t^{(3)}}-1)$-open edge is attached it. Then the choice of
$\t^{(2)}$ is possible from the set of cardinality less or equal
to $\theta(\xi_{\t^{(3)}})$. Since the arrival instant
$\xi(\t^{(2)})$ is open, the same concerns the choice of the
instant $\t^{(1)}$. Then the  number of possibilities to created
such a vertex is bounded  by $M_\theta^2$.

Now let us consider the second possibility when the
$(\xi_{\t^{(3)}}-1)$-open edge is attached not to
$w_{2s}(\xi_{\t^{(2)}})$ but to $w_{2s}(\xi_{\t^{(1)}})$. In this
case we have no restrictions for the choice of $\t^{(2)}$ but the
choice of $\t^{(1)}$ is restricted to the set of cardinality less
than $\theta(\xi_{\t^{(2)}})$. This gives the estimate of the
number of choices by $\t^{(3)}M_\theta$.

Regarding the cases (i)-(iii) and summing over all possible values
of $\t^{(3)}$, we conclude that in the case of triple
self-intersection the number of choices to construct a vertex with
one or two open arrival instants is bounded  by
$\max\left\{sM^2_\theta; s^2M_\theta/2\right\}\le s^2M_\theta$.

 Now it is not difficult to estimate the number of
possibilities to choose the values of $T^{3)}_s(\nu_3;r_3)$ with
given number $r_3$ of open arrival instants at the vertices of
triple self-intersections.

\v {\bf Lemma 3.4} {\it Given $\theta_{2s}$, the number of
possibilities $\vert T^{(3)}_s(\nu_3;r_3)\vert$ to construct
$\nu_3$ vertices of triple self-intersections with the total
number of $r_3$ open arrival instants is bounded by the following
expression;
$$
\vert T^{(3)}_s(\nu_3;r_3)\vert\le  {1\over \nu_3! }
{2\nu_3\choose r_3} \cdot \left(4s^2 \, M_\theta\right)^{\langle
r_3/2\rangle}\ \left({s^3\over 2!}\right)^{\nu_3-\langle
r_3/2\rangle} , \quad 0\le r_3\le 2\nu_3,
 \eqno (3.12)
$$
where we denoted
$$
\langle r_3/2\rangle = \cases{l, & if $r_3=2l$, \cr l+1, & if
$r_3=2l+1$\cr}.
$$
}
 \v {\it Proof.} First  we  point out $r_3$ arrival
 instants among $2\nu_3$ ones; this produces the factor
 ${2\nu_3\choose r_3}$.
Then we set the values of the last arrival $\tau$-instants
$\t^{(3)}_1<\dots< \t^{(3)}_{\nu_3}$.

The  vertex $\a'_j$ that has at least one open arrival instant can
be constructed by not more than $4sM_\theta$ choices of the values
of the arrival $\t$-instants $\t^{(1)}_j$ and $\t^{(2)}_j$, where
the factor $4$ estimates the choice of the head or the tail of the
open edges. The number of such vertices $\a'_j$ is not less than
$\langle r_3/2\rangle$.

The two first arrival instants $\t^{(1)}_i$ and $\t^{(2)}_i$ at
the vertices that have no open arrival instants can be chosen in
not more than $s^2/2$ ways.

Taking into account these observations and summing over the last
arrival $\tau$-instants, we   repeat the computations of (3.9) and
obtain (3.12). Lemma is proved. $\diamond$

\subsection{Counting the walks with BTS-instants}

In the previous subsections we have described the construction of
all walks with the same $\theta_{2s}$ that have closed and open
instants of simple and triple self-intersections. This is
important for the estimate of number of walks because the instants
of the open self-intersections leave a certain freedom  for the
walk to  continue its run at the non-marked instants of time. This
run can  differ from the canonical  run of the walk according to
the tree structure dictated by the corresponding $\theta_{2s}$.
So, the open self-intersection represent a potential possibility
to choose the canonical run or the non-canonical run at the
non-marked instants of time. The arrival instants followed by the
 non-canonical continuation  are the BTS-instants
considered in Section 2.

The BTS-instants are important when counting the number of
imported cells at one or another vertex of the walk. So, let us
inverse somehow the point of view of the previous subsections and
look at the set of walks that have a certain number $R$ of
BTS-instants. We assume that $R= \rho_2 + \rho_3 + \dots +
\rho_s$, where $\rho_k$ is the number of BTS-instants that happen
in the vertices $\b$ such that $\kappa(\b)=k$.

As usual, let us consider first the vertices of simple
self-intersections. Supposing that the walk $w_{2s}$ has $\rho_2$
BTS-instants, we conclude that there are at least $\rho_2$ open
instants of self-intersections in $w_{2s}$. Then this walk belongs
to the class $(\nu_2,r_2)$ with $r_2\ge \rho_2$.

If $w_{2s}$ is such that $\rho_3>0$, then there is a number of
vertices of triple self-intersections that contain at least one
open instant of self-intersection. Clearly, the number of such
vertices is greater than $\langle\rho_3/2\rangle$.

In what follows, we will need the estimates of choices of vertices
with $\kappa =2$ and $\kappa=3$ that have at least one open
instant of self-intersection. In contrast, for the vertices with
$\kappa\ge 4$ we take into account their number only. So, if
$w_{2s}$ is such that $\rho_k>0$ with $k\ge 4$, we observe that
this walk contains at least one vertex of $k$-fold
self-intersection.

 \v
At this point we terminate the study of such a rich and
interesting subject of even walks and pass to the estimates of
high moments of large random matrices.


\section{Estimates of the moments of random matrices}

In this section we use the results of our studies of even walks
and prove the estimates of the moments (1.2) in the asymptotic
regime $n\to\infty, s^3 = O(n^2)$. We are restricted to the
simplest case when the random matrix entries are given by bounded
random variables. Therefore our results represent a particular
case of statements formulated for the sub-gaussian random
variables  \cite{S} and random variables having a number of
moments finite \cite{R}. However, it should be stressed that the
arguments presented in \cite{R} and \cite{S} are not sufficient to
get  the full  proof of the estimates even in the simplest case
considered here. In the next subsection we address this question
in more details.

 Another important thing  to say is  that the results of
\cite{R,SS2,S} rely strongly on the assumption that the following
property of the Dyck paths is true
 \v
$$
B(\lambda) = \lim_{s\to\infty} {1\over C^{(s)}} \sum_{\theta_{2s}
\in \Theta_{2s}} \exp\left\{
 {\lambda M^{(s)}_\theta\over \sqrt s  }
\right\} <+\infty ,\quad \lambda>0, \eqno (4.1)
$$
where $\Theta_{2s}$ is the set of all Dyck paths of $2s$ steps,
$C^{(s)} = {\displaystyle (2s)!\over \displaystyle s!\, (s+1)!}$,
and $M^{(s)}_\theta = \max_{{0\le t\le 2s}} \theta_{2s}(t)$. This
estimate is closely related with the corresponding property of the
normalized Brownian excursion  that is proved
 to be true (see \cite{FJ} and references therein).
This is because the limiting distribution of the random variable
$M^{(s)}_\theta/\sqrt s$ considered on the probability space
generated by $\Theta_{2s}$ coincides with that given by this
half-plane Brownian excursion. We did not find any explicit
reference where (4.1) would be proved, but it is widely believed
to be true. So, we also prove our statements under the hypothesis
(4.1).

\subsection{Main estimate and the scheme of the proof}

The main result of the present section is given by the following
statement.

\V {\bf Theorem 4.1} {\it Consider the  ensemble of real symmetric
matrices $A_n$ (1.1) whose elements
$$
\left(A_n\right)_{ij} = {1\over \sqrt n}\,  a_{ij}
$$
are given by a family $\CA_n = \{ a_{i,j}, 1\le i\le j\le n\}$ of
jointly independent identically distributed random variables that
have symmetric distribution. We assume that there exists such a
constant $U\ge 1$ that $a_{i,j}$ are bounded with probability 1,
$$
\sup_{1\le i\le j\le n} \vert a_{ij}\vert \le U
 \eqno (4.2)
$$
and we denote the moments of $a$ by $V_{2m}$ with $V_2=1$;
$$
\E \{a_{ij} ^2 \}= 1, \quad \E \{a_{ij}^{2m} \}= V_{2m}.
 \eqno
 (4.3)
$$
If $s_n^3=  \mu  n^2$ with $\mu>0$, then  in the  $ n\to\infty$
the estimate }
$$
{\sqrt{\pi\mu} \over \ \, 4^{s_n}} \cdot \E \left\{\T \left(
A_n\right)^{2\lfloor s_n\rfloor} \right\}\le B(6\mu^{1/2})\,
\exp\{ C \mu\}
 \eqno (4.4)
$$
{\it is true with a constant $C$ that does not depend on $n$ and
on $V_{2m}, m\ge 2$. }

 \V Here we denoted by $\lfloor x\rfloor,
x>0$ the largest integer not greater than $x$. Not to overload the
formulas, we will omit this sign when no confusion can arise. Then
one can rewrite (4.4) in the following form similar to (1.5)
$$
{s_n!\, (s_n+1)!\over n(2s_n)!}   \cdot \E \left\{\T \left(
A_n\right)^{2 s_n} \right\}\le B(6\mu^{1/2})\,  \exp\{ C \mu\}
 \eqno (4.4')
$$
that is asymptotically equivalent to (4.4) due to the Stirling
formula.

 \v {\it Remarks.}

1.  We denote the limiting transition $s_n,n\to \infty$ such that
$s_n^3= \mu n^2$  by $(s_n,n)_\mu\to\infty$ or simply by
$(s,n)_\mu\to\infty$. It can be easily seen  from the proof of
Theorem 4.1 that the estimate (4.4) is also true in the limit
$s_n, n\to\infty$ such that $s_n^3/n^2\to\mu$. More generally, one
can prove (4.4) in the limit when $s_ n^3 = O(n^2)$ with $\mu =
\limsup_{n\to\infty} s^3_n/n^2\ge 0$.

2. Theorem 4.1 is also true for the ensemble of hermitian random
matrices. In this case the closed even walks that give non-zero
contribution to $\E\{\T A_n^{2s}\}$ are to be replaced by the
closed double-even walks. This could lower the estimate (4.4); in
particular, one could replace the coefficient $6$ in
$B(6\mu^{1/2})$ of (4.4).

3. One can compare these results with the non-asymptotic estimates
of the moments of real symmetric (or hermitian) random matrices of
the form (4.2) whose elements have joint Gaussian distribution.
Corresponding ensembles are represented by the Gaussian Orthogonal
Ensemble (GOE) and the Gaussian Unitary Ensemble (GUE) (see e.g.
\cite{M}).

It is proved in \cite{K3} that the moments of  GUE $m_{2k}^{(n)}$
are bounded by  (cf. (1.5))
$$
{n(2s)!\over s!\, (s+1)!}\left(1+  \a {k(k-1)(k+1)\over
n^2}\right)
$$
for all values of $k,n$ such that $k^3/n^2\le \chi$, where
$\a>(12-\chi)^{-1}$ with $\chi<12$. Then in this case the estimate
(4.4$'$) can be replaced by more explicit inequality, say
$$
{s!\, (s+1)!\over n (2s)!} m_{2s}^{(n)}\le \left(1+{\mu\over
11-\mu} \right), \quad \mu < 11.
$$
For more result on the non-asymptotic estimates of the moments of
random matrices, see e.g. \cite{L}.

 \v {\it Scheme of the proof of Theorem 4.1.}  We mainly
follow the scheme proposed  in paper  \cite{R} that slightly
modifies the approach of \cite{SS1,SS2,S} and gives more detailed
account on the computations involved.

The  strategy  is to consider the natural representation of
 $\T A_n^{2s}$ as the sum over the set $\CI_{2s}(n)$
of all possible  paths \mbox{$I_{2s}$ (cf. 2.1)} and split this
sum  into four sub-sums according to the properties of the graphs
of the paths  $I_{2s}\in \CI_{2s}(n)$.

These properties are related with the value of $\vert \bar
\nu_s\vert$ (3.2)
 and the maximal
exit degree of the graph $g(I_{2s})=g_{2s}$
$$
\D(I_{2s}) = \max_{\a \in \CV(g_{2s})} \deg_e(\a),
$$
where $ \deg_e(\a) =\vert \CD_e(\a)\vert$. Namely, we use the same
partition of $\T A_n^{2s}$ as in \cite{R,S} given by relation
$$
 \E \left\{\T  A_n^{2s} \right\} = \sum_{l=1}^4
Z_{2s}^{(l)},
 \eqno (4.5)
 $$
where
\begin{itemize}
\item $Z^{(1)}_{2s}$ is the sum over the set $\CI^{(1)}_{2s}\subset \CI_{2s}(n)$
of  all possible paths $I_{2s}$ such that
$\vert\bar\nu(I_{2s})\vert_1\le C_0s^2/n$ and there is no edges in
$\CE(I_{2s})$ passed by $I_{2s}$ more than two times;

\item $Z^{(2)}_{2s}$
is the sum over the set $\CI^{(2)}_{2s}$ of all the paths  $I_{2s}
$ such that $\vert\bar\nu(I_{2s})\vert_1\le C_0s^2/n$ and
$\D(I_{2s}) \le s^{1/2-\epsilon}$, $\epsilon>0$ and there exists
at least one edge of $\CE(I_{2s})$ passed by $I_{2s}$ more than
two times;

\item
$Z^{(3)}_{2s}$ is the sum over the set $\CI^{(3)}_{2s}$ of all the
paths $I_{2s} $ such that $\vert\bar \nu(I_{2s})\vert_1 \le C_0
s^2/n$ and $\D(I_{2s})  \ge s^{1/2-\epsilon}$;

\item $Z^{(4)}_{2s}$ is a sum over the set $\CI^{(4)}_{2s}$
of all the paths $I_{2s}$ such that $\vert\bar \nu(I_{2s})\vert_1
\ge C_0 s^2/n$.
\end{itemize}

The constant $C$ of (4.4) is chosen according the condition
$C>C_0+36$ where $C_0$ is determined in the proof of
 the estimate of $Z^{(4)}_{2s}$; it is sufficient to take
 $C_0>2eC_1^2 U^4$,
 where $C_1 = \sup_{ k\ge 1}{ \displaystyle 2k\over \displaystyle ((k-1)!)^{1/k}}$.
 The appropriate value of $\epsilon$ will be determined
 in the estimate of $Z^{(3)}_{2s}$; it is sufficient to choose
 $0<\epsilon <1/6$.

In the limit $(s,n)_\mu\to\infty$, the sub-sum $Z^{(1)}_{2s}$
contributes to (4.5) as a non-vanishing term, while other three sub-sums
are of the order $o(1)$. In the following four subsections we
consider these sub-sums one by one.
Let us explain the use of the results of Sections 2 and 3 in the proof of the estimates of $Z^{(i)}_{2s}$.
\v
In Section 3 we presented the three types of estimates; first we estimated

A) the number of walks with given $\theta_{2s}$ and $\bar \nu_s$;

\noindent then we specified these estimates and studied

B) the number of walks with a number of simple open self-intersections;

\noindent finally, we estimated

C) the number of walks that have a number of triple self-intersections with open
arrival instants.
\v
\noindent In the present section we use these estimates
completed in some cases by the information about

D) walks with self-intersections
that produce factors $V_{2m}, m\ge 2$.
\v

\noindent Namely,
\begin{itemize}
\item to estimate $Z^{(1)}_{2s}$, we use mostly parts (A) and (B) described above;
here our computations repeat almost word-by-word those of
\cite{SS2,S};

\item  to estimate $ Z^{(2)}_{2s}$, we use the parts (A), (B), and
(D); the structure of simple and triple self-intersection that
produce the factors $V_{2m}$ is studied in more details than it is
done in \cite{R,S};

\item  to estimate $Z^{(3)}_{2s}$, we use the items (A), (B), (C), and  (D);
this is the most complicated part of the present  section that
involves the results of Section 2; here we use the new ingredient
of the structure of even closed walk  we called the primary and
imported cells; this makes our arguments and computations
essentially different from those of \cite{S} and \cite{R};

\item  to estimate $Z^{(4)}_{2s}$, the part (A) is sufficient;
here we slightly modify the reasoning of \cite{R} by adding some
missing elements of the proof.

\end{itemize}
Let us start to perform the program presented.

\subsection{Estimate of $Z^{(1)}_{2s}$}

Taking into account observations and   results of Section 3, we
can write that
$$
Z^{(1)}_{2s} \le \sum_{\theta\in \Theta_{2s}} \sum_{\sigma=0}^{C_0
s^2/n} \sum_{ \bar \nu :\ \vert \vec \nu\vert_1 = \sigma}
{n(n-1)\cdots (n-\vert \CV(g_{2s})\vert+1 )\over n^s} \times
$$
$$
\sum_{r=0}^{\nu_2}\vert T^{(2)}_s(\nu_2;r;\theta_{2s })\vert \cdot
\prod_{k=3}^s \vert T^{(k)}{(\nu_k;\theta_{2s})}\vert \cdot
W_s(\bar \nu_s; r),
 \eqno (4.6)
$$
where $\vert T^{(2)}_s(\nu_2;r)\vert$ represents an estimate of
the number of possibilities to point out  $\nu_2$ vertices of
simple self-intersections such that  $ r$ self-intersections are
non-closed (3.11), the variables $\vert
T^{(k)}{(\nu_k;\theta_{2s})}\vert$ with $k\ge 3$ are as in (3.6)
and $W_s(\bar \nu_s; r)$ is given by (3.3).

Remembering that $\vert \CV(g_{2s})\vert = s+1-\vert \bar \nu\vert_1$
and that $\s= \sum_{k=2}^s (k-1)\nu_k$, we can write the following
inequality
$$
n {(n-1)\cdots (n-s+\s)\over n^{s-\s}} \cdot {1\over n^{\s}} \le
n\exp\left\{ -{s^2\over 2n} + {s\s\over n} \right\}   \cdot
\prod_{k=2}^s {1\over n^{(k-1)\nu_k}}\ .
 \eqno (4.7)
$$
In this computation we have used the following elementary result.

\V {\bf Lemma 4.1} (\cite{SS1}). {\it If $s<n$, then  for any
positive natural $\s$ the following estimate holds}
$$
\prod_{k=1}^{s-\s} \left(1 - {k\over n}\right) \le
\exp\left\{-{s^2\over 2n}\right\} \exp\left\{{s\s\over n}\right\}.
 \eqno (4.8)
$$

\v

{\it Proof.} The proof mainly repeats the one of \cite{SS1}. We
present it for completeness. Elementary computations show that
$$
\prod_{k=1}^{s-\s} \left(1 - {k\over n}\right) = \exp\left\{
\sum_{k=1}^{s-\sigma} \log \left(1-{k\over n} \right) \right\} =
\exp\left\{ - \sum_{k=1}^{s-\sigma} \left( \sum_{j=1}^\infty
{k^j\over j n^j} \right)\right\}
$$
$$
\le \exp\left\{ - \sum_{k=1}^{s-\s} {k\over n}\right\} \le
\exp\left\{ -{(s-\sigma)^2\over 2n} \right\} \le \exp\left\{
-{s^2\over 2n}+ {s\sigma\over n}\right\}.
$$
Lemma is proved. $ \diamond$

\V

Taking into account results (3.6) and (3.11)  of Lemmas 3.2 and
Lemma 3.3 and using the estimate of $W_n$ (3.3),
we derive from (4.6) with the help of (4.7)  the following inequality
$$
Z^{(1)}_{2s}\le
n\\
\exp\left\{C_0{s^3\over n^2}  -{s^2\over 2n}\right\}
\sum_{\theta\in \Theta_{2s}}\ \sum_{\sigma= 0}^{C_0 s^2/n} \ \
\sum_{ \bar \nu :\ \vert \bar \nu\vert_1 = \sigma} {1\over
\nu_2!}\left( {s^2\over 2n} + {6 s M_\theta\over n}\right)^{\nu_2}
$$
$$
\times  \prod_{k=3}^s {1\over \nu_k!} \left( {(2k)^k \ s^k\over
(k-1)!\ n^{k-1}}\right)^{\nu_k}.
$$
 Passing to the sum over $\nu_i\ge 0, i\ge 2$ without any restriction,   we can write that
$$
Z^{(1)}_{2s}\le n \, {\exp\{C_0\mu \} } \sum_{\theta\in
\Theta_{2s}}\exp\left\{{6 M_\theta\sqrt{\mu}\over \sqrt s}+36\mu
+\sum_{k\ge 4}{(C_1 s)^k\over n^{k-1}}\right\},
 \eqno (4.10)
$$
where $C_1 = \sup_{k\ge 3} {2k/ ((k-1)!) ^{1/k}}$. It is easy to
see that   the Stirling formula implies relation
$$
n\vert \Theta_{2s}\vert = n C^{(s)}= {n (2s)! \over  s! \ (s+1)! }
= { 4^s\over \sqrt{\pi \mu}} \ (1+o(1)),\quad (s,n)_\mu\to\infty.
$$
 Multiplying
and dividing the right-hand side of (4.10) by $C^{(s)}$ and using
(4.1), we conclude that
$$
{\sqrt{\pi \mu}\over 4^s} \cdot
 Z^{(1)}_{2s}
\le  B(6 \mu^{1/2})\cdot \exp\left\{C\mu\right\} , \quad
(s,n)_\mu \to\infty, \eqno (4.11)
$$
where $C>C_0+36$. This is the estimate for $Z^{(1)}_{2s}$ we need.


\subsection{Estimate of $Z^{(2)}_{2s}$}

In the present subsection we study the walks whose weight contains
at least one factor $V_{2m}$ with $m\ge 2$. The first important
observation here is that it is sufficient to consider in details
the multiple edges attached to the vertices of the
self-intersection degree $\kappa\le 3$ only. The contribution of
the multiple edges attached to other vertices will be replaced by
the bound $U$ with the appropriate degree.

Also we have  to say  that here one meets the following general
inconvenience of the method: the classes of equivalence of the
paths and the walks are characterized by the set $\bar \nu$ that
determines the multiplicities of the vertices of
self-intersections. However, this description does not take
explicitly  into account the multiplicities of the edges.
Sometimes this makes the  study of the classes of walks whose
weight contains the factors $V_{2m}, m\ge 2$ rather cumbersome.

\subsubsection{Simple self-intersections of the first kind}

Let us consider first the vertices attached to the edges that
produce the factors $V_4$. Assume that among $\nu_2$ vertices of
simple self-intersections
 there are
$r_2$ open vertices of
 self-intersection.
We assume also that among $\nu_2 - r_2$ vertices of simple
self-intersections there are $q_2$ vertices $\b$ such the
following condition is verified: the second arrival to $\beta$ at
the marked instant is performed along the edge oriented in the
same direction as the edge corresponding to the first arrival to
$\beta$. We denote the  edge of this first arrival by $(\a,\b)$.

\begin{figure}[htbp]
\centerline{\includegraphics[width=9cm]{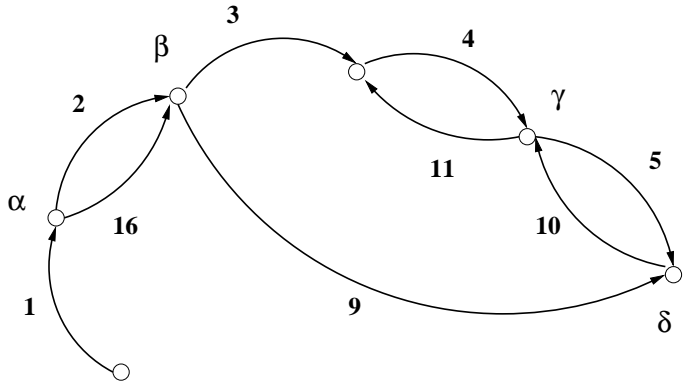}}
\caption{\footnotesize{The marked part of  $g(w_{18})$ with the
simple self-intersections of  different types}}
\end{figure}
\v
 \noindent We say that these $q_2$ simple self-intersections are
of the type one or I-type simple self-intersections. On Figure 3
we present the marked edges of the graph of the walk that has one
simple  self-intersection $(t',t'')=(2,16)$  of the I-type.

\V

Let us formulate the following elementary proposition.

\V {\bf Lemma 4.2. (\cite{SS1})} {\it Given the position of the
second arrival instant
 $\t''$ of simple
self-intersection of the first kind,  there exist not more than
$\D^{(\a)}$ possibilities to choose the first arrival $\t$-instant
$\t'$, where
$$
\D^{(\a)} = \deg_e(\a), \quad \a = w_{2s}(\xi_{\t''} -1).
$$
 }

 {\it Proof. } The proof  immediately follows from  the
observation that the vertex of the simple I-type self-intersection
$(\t',\t'')$ is such that the edge of first arrival
 belongs
to the exit cluster  $\b\in \CD_e(\a)$ of the sub-walk
$w_{[1,\xi_{\t''}]}$ of $w_{2s}$. $\diamond$

\V\v {\bf Corollary of Lemma 4.2. } {\it If one is restricted with
the class of walks with given $\bar \nu$ and  bounded maximal exit
degree
$$
\D(I_{2s}) = \max_{\a\in \CV(g_{2s}) }
\deg_e (\a)\le d,
$$
 then the number of possibilities to
construct such a walk of $r_2$ open self-intersections with $q_2$
of the  I-type self-intersections
 is
bounded by
$$
\vert T(\nu_2,r_2,q_2) \vert \cdot \prod_{k=3}^s \vert T^{(k)}_s(\nu_k)\vert \cdot
 W_s(\bar \nu;r),
 \eqno (4.12)
$$
where
$$
\vert T(\nu_2;r_2;q_2)\vert  \le {(2sM_\theta)^{r_2} \
 (sd)^{q_2}\over r_2!\,q_2!\, (\nu_2-r_2-q_2)! } \cdot
 \left( {s^2\over 2}\right)^{\nu_2-r_2-q_2}.
 \eqno (4.13)
$$}
\v {\it Proof.} Relation (4.13) corresponds to the choice of the
vertices of simple self-intersections. Then (4.12) obviously
follows after the reasoning similar to that used in
 \mbox{Section 3.}

\subsubsection{Simple self-intersections of the second kind and chains of them}

 The case of simple self-intersections of the first kind is
described in \cite{R,SS1}. Now let us consider other possibilities
to construct the walks whose weight  contains the  factors $V_4$.

First let us assume that among remaining $\nu_2 - r - q_2$
vertices there are $p_2$ vertices $\g$ that verify the following
property: the second arrival at $\g$ at the marked instant is
performed along the edge $\tilde e = (\d,\g)$ such that the marked
edge $(\g,\d) = \hat e$ already exists in $g(w_{2s})$. We say that
these simple self-intersections are of the type two, or II-type
simple self-intersections. On Figure 4 we give an example of the
walk $w_{2s}$ with $p_2=2$ and $q=1$. We show there the marked
edges of the graph $g(w_{2s})$.

To construct the self-intersection of II-type, the walk has to
close the edge $\hat e$. The direction of this closure indicates
two different ways to create the II-type self-intersection.

In the first case the closure of $\hat e = (\g,\d)$ is performed
in the inverse direction $(\d,\g)$. Then the marked  edge
 $\tilde e$ can be created after the marked arrival at $\d$. We
 see that the vertex $\d$ is by itself the vertex of a
 self-intersection. When arrived at $\d$, the walk has to decide
 about the choice of the next vertex; there is not more than three
 marked edges that arrive at $\g$ and the number of choices of the
 vertex $\g$ is bounded by $3$.
  Let us recall that we consider
the edges attached to vertices with $\kappa\le 3$ only. Therefore
the simple
 self-intersection enters into the sub-sum we study with the
 factor of the order $Const\cdot  {s^2\over 2n}\cdot  {3\over n}= O(s^2/n^2)$.

Another possibility is given by the closure of the edge
 $\hat e = (\g,\d)$ in the direction $(\g,\d)$. In this case the
 walk has to perform an open self-intersection before this
 closure and such a II-type self-intersection contributes by the
 factor of the order $M_\theta/n$, where $M_\theta$ estimates the
 number of choices of $\t''$. We do not give the  rigorous proofs of
 the  statements presented above.

Now let us consider the situation when  $p_2$ vertices of simple
self-intersections of the second type are organized into several
chains  of neighboring self-intersections as it is shown on Figure
4. Regarding one group only that contains $l_1$ elements in the
chain, we see that there is less than $3s^2$ possibilities to
produce the last self-intersection at the vertex $\g^{(1)}$ where
$w_{2s}$ arrives at the second time at the instant $t$. To produce
the second element of this chain, the next in turn instant of
self-intersection is to be chosen from the exit cluster of $w(t)$.
Therefore the number of possibilities to perform this is not
greater than $\sup_t \deg_e(w(t))= \D(I_{2s}) = \D$. Then the
total number of possibilities to produce such a chain is bounded
by $3s^2 \D^{l_1-1}\le s^2 \D^{l_1}$, where we assumed for
simplicity that $\D\ge 3$.

If there are $v$ chains of $l_i$ elements with $l_1+\dots+l_v=p$,
then the number of possibilities is bounded by $s^{2v}\D^p$ and
the number of vertices in the corresponding graph $g(w_{2s})$ is
bounded by $s- \vert \vec \nu\vert_1 -v $. This means that such a
configuration enters into the sum with the factor
$$
{1\over v!}  \left( {s^2\over n}\right)^v \sum_{l_1+\dots+l_v=p-v}
\left( {\D\over n}\right)^{l_1} \cdots  \left( {\D\over
n}\right)^{l_1} \le {1\over v!}  \left( {s^2\over n}\right)^v
\left(\sum_{l_1\ge 1} \left( {\D\over n}\right)^{l_1}\right)^v \le
{1\over v!}  \left( {2 s^2\D\over n^2}\right)^v.
$$
Here we have taken into account that $\D/n\le s^{1/2}/n = o(1)$ as
$n\to\infty$.

\subsubsection {Triple self-intersections and factors $V_{2j},\, j\ge 2$}

Let us consider first the vertices of the triple
self-intersections that can be seen at the edges that produce the
factor $V_4$. Regarding the simple self-intersections of the type
one, we see that one can add an arrival edge at $\b$ and keep the
factor $V_4$ with no changes. If there are $q_3$ vertices $\b$
with $\kappa(\b)=3$ of this type, then there is not more than
$(s^2\D)^{q_3}/q_ 3!$ possibilities to choose the instants to
create such a group. This group enters $Z^{(2)}_{2s}$ with the
factor $n^{-2q_3}$. Some of the vertices of the chains described
above can be also the vertices of triple self-intersection. Then
each of the factors $(\D/ n)^{l_i}$ given above should be replaced
by
$$
\left( {\D\over n}\right)^{l_i}\sum_{u_i=0}^{l_i} {l_i+1\choose
u_i}\left({s\over n}\right)^{u_i}= \left( {\D\over
n}\right)^{l_i}\left(1 + {s\over n}\right)^{l_i} = \left( {\D\over
n}(1+o(1))\right)^{l_i}.
$$
The same concerns each of the $v$ factors $s^2/n$.

\v

Let us pass to factors $V_6$. To consider these, we have to study
the vertices of triple self-intersections or the mixed cases given
by the vertices of simple self-intersections of type two and/or
the vertices of triple self-intersections. Slightly modifying the
previous reasonings, it is easy to see that $(V_6)^Q$ enters  with
the factor bounded in these two cases by
$$
{1\over Q!} \left( {s \D^2\over n^2} + {s\D^3\over
n^3}(1+o(1))\right)^Q.
$$
The first factor takes into account either  the edges whose ends
are the vertices of simple self-intersections of the types I and
II or the triple self-intersections of the type one; the second
one corresponds to the triple self-intersection of the type two
and the chains of them. Both of these types an be determined by
straight analogy with the types of simple self-intersections. We
do not present the details here.

Taking into account $q_3$ vertices of triple self-intersections
described above and assuming that
 $v$ chains of edges are constructed with the help of $p_3$ vertices
 of triple self-intersections,
we can write that the contribution of the vertices of triple
self-intersections that produce moments $V_{2j}$ with $j\ge 2$ is
given by the factor bounded by
$$
{1\over (q_3 + p_3 + Q)!} \left( {s^2\D\over n^2} V_8 +
{s\D^2\over n^2} V_{10} (1+o(1))+ {s\D^3\over
n^3}V_{10}\right)^{q_3+p_3+Q}.
$$
Here the factor $1+o(1)$ corresponds to the chains of vertices of
triple self-intersections of the type two.

\v The factors $V_8$ can arise due to the presence of triple
self-intersections of the special type similar to the type two of
the simple and triple self-intersections. It is not hard to see
that the factor $(V_8)^P$ enters together with the number
estimated by expression ${1\over P!} (s\D^4/n^4)^P$.

\v

Gathering the observations of this subsection and using the
computations of the previous subsection, we conclude that
$$
Z^{(2)}_{2s} \le n \sum_{\theta\in \Theta_{2s}}
\sum_{\s=0}^{C_0s^2/n} \ \sum_{\bar \nu:\  \vert \bar
\nu\vert_1=\s} \ \sum_{r_2=0}^{\nu_2} \ \sum_{
p_2+q_2=0}^{\nu_2-r_2} \ \sum_{Q_3+P+\nu_3'=\nu_3}\
\exp\left\{-{s^2\over 2n} + C_0 \mu\right\} \times
$$
$$
 {1\over (\nu_2 - p_2-q_2-r_2)!}
\left({s^2\over 2n}\right)^{\nu_2 - p_2 - q_2- r_2} \cdot {1\over
r_2!} \left({3s M_{\theta}\over n}\right)^{r_2}\cdot {1\over q_2!}
\left({s\D \over n}V_4\right)^{q_2}\times
$$
$$
{1\over p_2!}\sum_{v=1}^{p_2} {1\over v!} \left( {2s^2\D\over n^2
}V_4 \right)^v\cdot {1\over Q_3!} \left( {s^2\D\over n^2} V_4 +
{2s\D^2\over n^2} V_6 + {s\D^3\over n^4}V_6\right)^{Q_3} \times
$$
$$
{1\over P!} \left( {s^3\over n^4}\right)^P V_8^P \cdot
I_{[1,s]}(p_2+q_2+P+Q_3)\cdot{1\over \nu'_3!}\left({36 s^3 \over \
n^2}\right)^{\nu'_3}\cdot {1\over \nu_k!}\prod_{k=4}^s
\left({C_1^k U^{2k} s^k\over n^{k-1}}\right)^{\nu_k},
 \eqno (4.14)
$$
where we have replaced all factors $1+o(1)$ by $2$ and have
denoted by $I_B(\cdot)$ the indicator function
$$
I_ B(x)  =\cases{ 1, & if $x\in B$,\cr 0, & if $x\notin B$\cr}
$$
and by $\nu'_3$ the number of vertices from $\CN_3$ that are not
included into the subsets considered above. Although we could use
inequality $V_{2m}\le U^{2m}$ in (4.14), we prefer to keep
 the factors $V_{2m}$
to indicate clearly the origin of the corresponding factors.

Remembering that $\D\le s^{1/2-\epsilon}$ with $\epsilon>0$, we
repeat the computations of the previous subsection that lead to
(4.10) and  deduce from (4.14)  inequality
$$
{\sqrt {\pi \mu} \over 4^s} Z^{(2)}_{2s} \le \exp\{(C_0+36) \mu\}
\cdot B (6 \mu^{1/2}) \cdot \left( \exp\left\{ {\sqrt{\mu }\over
s^\epsilon}\, V_4(1+o(1))
 \right\} - 1 \right).
$$
 Then
$$
{\sqrt {\pi \mu} \over 4^s} Z^{(2)}_{2s} = o(1), \quad \hbox{as }
(s,n)_\mu\to\infty.
 \eqno (4.15)
$$
This estimate shows that the paths $\CI^{(2)}_{2s}$ do not
contribute to $m_{2s}^{(n)}$ in the limit we
consider.


\subsection{Estimate of $Z^{(3)}$ }

We have seen in the previous subsection that the presence of
$V_{2m}$ in the weight is related mainly with the I-type simple
self-intersections  and therefore with the exit degree of a
vertex. The exit degree of $\b$ is determined as the cardinality
of the exit cluster $\CD_e(\b)$ defined in subsection 2.2. In the
present subsection we concentrate on  the classes of walks such
that their  maximal exit degree is large; $\D =\max_\b \vert
\CD_e(\b)\vert \ge s^{1/2-\epsilon}$.

\v

Given $\theta_{2s}\in \Theta_{2s}$, we determine the canonical
walk $w^{(0)}_{2s} = w(\theta_{2s})$ as the walk without
self-intersections constructed with the help of $\theta_{2s}$.
Clearly, the graph $g(w^{(0)}_{2s})$ represents a rooted
half-plane tree of $s$ edges
 $\CT_s =
\CT(w^{(0)}_{2s}) = \CT(\theta_{2s})$ introduced in Section 1. We
determine the vertices and the exit clusters of the tree $\CT_s$
in the obvious way.

\v

\subsubsection{Exit sub-clusters and $\CL$-property of the Dyck paths}

Given a walk $w_{2s}$, we consider a vertex $\b$ of $g(w_{2s})$
and denote by $0< \zeta_1< \dots <\zeta_L<2s$  the arrival
instants at $\b$ that represent either  primary or imported cells.
We determine a partition of the exit cluster $\CD_e(\b)$ into
subsets $\CD^{(l)}(\b), 1\le l\le L$, where the elements of
$\CD^{(l)}_\b$ are given by the marked edges created during the
time interval $(\zeta_l, \zeta_{l+1})$ with $\zeta_{L+1}\equiv
2s$.
 If there is no such marked edges, we  say that
the corresponding subset  $\CD^{(l)}(\b)$ is empty. The following
statement is a simple consequence of the definition of the primary
and imported cells.

\v {\bf Lemma 4.3} {\it Consider a walk $w_{2s}$ and its Dyck path
$\theta(w_{2s})$ with the corresponding tree $\CT(\theta)$. Then
the edges of the same subset $\CD^{(l)}(\b)$ correspond the edges
of the tree $\CT(\theta)$ that belong to the same exit cluster of
 $\CT(\theta)$. Denoting by $L'$ the number of all such exit clusters of $\CT(\theta)$
 that
correspond to $\CD^{(l)}(\b), 1\le l\le L$, we have
$$
L'\le L.
 \eqno (4.16)
$$
}

{\it Proof.} Let us denote by $ t^{(l)}_1=t_1$ and $
t^{(l)}_2=t_2$ the instants when the first and the last elements
of $\CD^{(l)}(\b)$ are created. According to the definition of
$\zeta_i$, the time interval $[\zeta_{l}+1,\zeta_{l+1}-1]$
contains the non-marked arrival instants  at  $\b$ only, if they
exist, and these arrivals do not represent the imported cells.
Then the sub-walk $\tilde W= w_{[ t_1,  t_2-1]}$ starts and ends
at $\b$ and is of the tree-type structure.   This means that after
a series of reductions described in subsection 2.3 this sub-walk
can be reduced to the empty sub-walk. Not to overload this paper,
we do not present corresponding rigorous definitions and proofs.

Since $\tilde W$ is of the tree-type structure, then  the edges of
$\CD_{\b}^{(l)}$ correspond to the children of the same parent in
$\CT(\theta)$. It can happen that the edges of $\CT(\theta)$ that
correspond to different clusters $\CD_{\b}^{(l)}$ and
$\CD_{\b}^{(l')}$ have the same parent. Lemma 4.3 is proved.
$\diamond$

\V Now let us introduce an important characterization of the Dyck
paths that represent a simplified version of  the property
proposed in \cite{S} and used in \cite{R}.

 \v \v {\bf Definition
4.1}. {\it
 We say that the Dyck
path $\theta\in \Theta_{2s}$ verifies the $\CL(m)$-property, if
there exists $\a\in \CV(\CT(\theta_{2s}))$ such that
$\deg_e(\a)\ge m$. We denote by $\Theta_{2s}^{(m)}$ the subset of
Dyck paths that verify this property.}

\V Let us explain the use of the $\CL$-property in the estimate of
$Z^{(3)}_{2s}$.  The set of walks involved in $Z^{(3)}_{2s}$ is
characterized by the fact that the graph of each of these walks
contains at least one vertex $\b$ such that its exit degree
$D_e(\b)$ is greater than $d\ge s^{1/2-\epsilon}$. We assume
$\b_0$ to be the first vertex of this kind in the chronological
order. Let us denote by $N$ the self-intersections degree of
$\b_0$ and by $K$ the total number of  the BTS-instants performed
by the walk. Then one can observe that some of the Dyck paths
$\theta$ cannot be used to construct the walks of the type
$(d,N,K)$ determined.

Indeed, it follows from the corollary of Lemma 2.1 that  the
number $L$ of primary and imported cells at $\b$ is bounded by
$2N+K$. According to (4.16), the number of parents $L'= L'_\b$ in
the corresponding tree $\CT$ is also bounded by $2N+K$. If
 $\theta'\notin \Theta_{2s}^{(d/(2N+K))}$, then
any tree $\CT(\theta')$ has no vertices with the exit degree
greater than or equal to $ d/(2N+K)$. Then obviously
$\deg_e(\b_0)$ is strictly less than $d$.

\v In \cite{S} it is argued that the subset of Dyck paths
$\Theta_{2s}^{(m)}$ has an exponentially  bounded cardinality with
respect to $C^{(s)} = \vert \Theta_{2s}\vert$;
$$
\vert \Theta_{2s}^{(m)}\vert \le a s^b C^{(s)} \exp\{ -C_2m\},
\eqno (4.17)
$$
where $a=1, b=2$ and $C_2 $ is a  constant. In \cite{K} (4.17) is
proved with $a=2, b=1$ and $C_2 = \log (4/3)$.

\subsubsection{General expression to estimate $Z^{(3)}_{2s}$}

To estimate the sum $Z^{(3)}$, we determine the values of
variables $d,N,K$ and consider the paths such there exists a
vertex $\b_0$ of the corresponding graph of the exit degree
deg$_e(\b_0)=d$ with the self-intersection degree $\kappa(\b_0) =
N$; the sum over $\theta$ is restricted to the subset
$\Theta_{2s}^{(d/(2N+K))}$ and according to Lemma 2.1, the
structure of the path is such that the corresponding walk has $R$
open instants of self-intersection with $R\ge K$ and
$R=\rho_2+\dots +\rho_s$, where $\rho_k$ is the number of open
 arrival instants at the vertices of self-intersection
degree $k$.

Repeating the reasoning of the  subsection 4.3 leading to the
estimate (4.16) and taking into account the arguments of the
previous subsection, we conclude that $Z^{(3)}_{2s}$ is bounded by
the following expression
$$
Z^{(3)}_{2s}\le n \exp\left\{ C_0 {s^3\over n^2} - {s^2\over
2n}\right\} \ \sum_{\theta\in \Theta_{2s}} \  \sum_{d\ge
s^{1/2-\epsilon}}\ \ \sum_{N=1}^s \ \ \sum_{K\ge 0}\ \
 \ I_ {\Theta^{(d/(2N+K))}_{2s}}(\theta)
 $$
 $$
 \times  \sum_{\s = 0}^{C_0 s^2/n} \ \ \sum_{\bar
\nu:\, \vert \bar \nu\vert_1=\s}^{(N)}\ \  \sum_{R= K}^\s \ \
\sum_{\stackrel{r_2+\dots +r_s=R}{r_i\ge 0}}\  \
 \ {1\over r_2!}\cdot \left(
{6sM_\theta\over n}\right)^{r_2}
$$
$$
\times {1\over (\nu_2-r_2)!} \left( {s^2\over 2n} + {sdV_4\over
n}(1+o(1)) \right)^{\nu_2-r_2} \cdot {1\over (\nu_3-\langle
r_3/2\rangle )!} \left( {36 V_{12} s^3\over n^2}\right)^{\nu_3 -
\langle r_3/2\rangle}
$$
$$
\times {1\over \langle r_3/ 2\rangle!} \left( {72 V_{12}s^2
M_\theta\over n^2}\right)^{\langle r_3/2\rangle} \cdot \
\prod_{k=4}^s\ \ {(k-1)\nu_k\choose r_k}
 \cdot {1\over \nu_k!} \left( {C_1^k
U^{2k} s^k\over n^{k-1}}\right)^{\nu_k}.
 \eqno (4.18)
$$
In this expression
 the sum $\sum^{(N)}_{\bar \nu}$ is taken over the sets $\bar
\nu $ such that $\nu_N\ge 1$. The term ${(k-1)\nu_k\choose r_k}$
stands for the choice of $r_k$ BTS-instants among $(k-1)\nu_k$
arrival instants at the corresponding vertices; here we assume
that ${0\choose l}=\delta_{l,0}$ for any integer $l\ge 0$.  The
factor $72=2\cdot 36$ gives the estimate of the corresponding
choices for the vertices of triple self-intersections multiplied
by the estimate that comes from the corresponding part of $W_n$.

Let us explain the presence of the factors $V_{12}$ in (4.18). The
last product corresponds to the vertices with $\kappa\ge 4$.
Regarding the vertex $\b$ of the triple self-intersection, we
denote three marked arrival edges at $\b$ by $(\a_i,\b)$,
$i=1,2,3$. The vertices $\a_i$ can coincide between them. However,
we see that the vertex $\b$ produces that maximal weight in the
case when all $\a_i$ are distinct and each of $\a_i$ is the head
of two  marked edges $(\b,\a_i)$. This gives the factor
 $V_4^3\le V_{12}$. Here we have taken into account the agreement
 that $\kappa(\a)\ge 1$, $\a\in \CV(g_{2s})$.

 Remembering that $s = \mu^{1/3} n ^{2/3}$, we can write
equality
$$
U^{2k} C_1^k {s^{k-1} s \over n^{k-1}} = U^{2k} C_1^k \mu^{k/3}
{n^{2/3}\over n^{(k-1)/3} } = \d_n H_n^{k-4} \cdot {1\over
n^{(k-1)/12}},
$$
where
$$
\d_n = {U^8C_1^4\mu^{4/3} n^{-1/12}}\quad {\hbox{and}}\quad H_n =
{U^2 C_1 \mu^{1/3} n^{-1/12}} .
 \eqno (4.19a)
$$
Summing a part of (4.18) over all possible values of
 $r_i, i\ge 4$, we obtain with the help of multinomial theorem that
$$
\sum_{r_4+\dots +r_s=\rho_4} \ \prod_{k=4}^s \ {1\over r_k!}
\left( (k-1)\nu_k\right)^{r_k} \cdot {1\over \nu_k!} \left( {C_1^k
U^{2k} s^k\over n^{k-1}}\right)^{\nu_k}
$$
$$
= {\s_4^{\rho_4}\over \rho_4!}\cdot {1\over n^{\s_4/12}} \
\prod_{k=4}^s {\d_n^{\nu_k}\over \nu_k!} \cdot H_n^{(k-4)\nu_k},
 \eqno (4.19b)
$$
where we denoted
 $\s_4 = \sum_{k\ge 4}(k-1) \nu_k$.
Here we have used the obvious estimate
$$
{(k-1)\nu_k\choose r_k}\le  {1\over r_k!} \left(
(k-1)\nu_k\right)^{r_k}.
$$

 Now we are ready to
estimate the right-hand side of (4.18). To better explain the
principle and the  estimates, we split our considerations into
three parts. In the first one we consider sub-sum given by the
right-hand side of (4.18) with $\s_4=0$ and  $r_3=0$. We denote
this sub-sum by $\check Z^{(3)}_{2s}$. When estimating
 $\check Z^{(3)}_{2s}$, we illustrate the main tools of the present
subsection.

Then we denote the sub-sum of (4.18) with $\s_4=0$ and $r_3\ge 0$
by $\tilde Z^{(3)}_{2s}$ and the sub-sum of (4.18) with $\s_4\ge
1$ by $\breve Z^{(3)}_{2s}$. Obviously,
$$
Z^{(3)}_{2s}\le \tilde Z^{(3)}_{2s} + \breve Z^{(3)}_{2s}.
$$
 We show that $\tilde  Z^{(3)}_{2s}=o(1)$ and $\breve
 Z^{(3)}_{2s}=o(1)$ as $(s,n)_\mu\to\infty$ in the second and the third parts of the
 proof. This implies the conclusion that $ Z^{(3)}_{2s}=o(1)$ as
 $(s,n)_\mu\to\infty$.

\subsubsection{The basic case of $\s_4=0$ and $r_3=0$}

Relation $\s_4=0$ implies equality $\rho_4=0$. Also we observe
that the sum over $N$ runs from $1$ to $3$. Remembering that the
corresponding sum is denoted by $\check Z^{(3)}$
 and taking into account that $r_2=R$, we can
write the following estimate
$$
\check Z^{(3)} \le  \sum_{d\ge s^{1/2-\epsilon}} \
\sum_{\s=0}^{C_0s^2/n} \ \sum_{\nu_2+2\nu_3 = \s}\
 \sum_{K=0}^\s \ \
 \sum_{r_2=K}^{\nu_2}\ \sum_{\theta\in\Theta_{2s}}\
   {1\over r_2!}\cdot \left( {6sM_\theta\over
n}\right)^{r_2}
 \cdot {1\over (\nu_2-r_2)!} \left( {s^2\over 2n}
\right)^{\nu_2-r_2}
$$
$$
\times {1\over \nu_3!}  \left( {36 V_{12} s^3\over
n^2}\right)^{\nu_3}\cdot  3s \cdot \exp\left\{{C_0s^3\over n^2}
-{s^2\over 2n}\right\} \cdot \exp\left\{ {2sd\over n} V_4 - {C_2
d\over 6+K}\right\}\cdot I_{\Theta_{2s}^{(d/6+K)}}(\theta).
 \eqno (4.20)
$$

Let us denote $X= s^2/(2n)$ and $\Phi = 6s M_\theta/n$ and
consider  the sum over $r_2$;
$$
S_{\nu_2}^{(K)}(X,\Phi) =  \sum_{r_2=K}^{\nu_2}
{1\over(\nu_2-r_2)!} X^{\nu_2-r_2}\cdot {1\over r_2!} \Phi^{r_2} =
{1\over\nu_2!}\sum_{r_2=K}^{\nu_2} {\nu_2\choose r_2}
X^{\nu_2-r_2} \Phi^{r_2}.
$$
Multiplying and dividing by $h^K$, we conclude that if  $h>1$,
then
$$
S_{\nu_2}^{(K)}(X,\Phi) = {1\over h^K \nu!} \left( \Phi^{\nu_2}
h^K + \dots  + {\nu_2\choose K} X^{\nu_2-K} \Phi^K h^K\right) \le
{1\over h^K}  \cdot {(X+ h\Phi )^{\nu_2}\over \nu_2!}.
 \eqno (4.21)
$$

This inequality illustrates the  principle we use to estimate
$\check Z^{(3)}$. Regarding the last line of (4.20), we see the
factor $\exp\{-s^2/(2n)\}$ that normalises the sum of the
 powers of $s^2/n$ diverging in the limit
$(s,n)_\mu\to\infty$. This sum is in certain sense  not complete
because of the presence of powers of asymptotically bounded
factors $6sM_\theta/n$, and this makes possible to use the
corresponding exponentially decaying factor $h^{-K}$.

Regarding the normalized sum over $\Theta_{2s}$ as a kind of the
the mathematical expectation $E\{\cdot\}$, we use  elementary
inequality $E\{ fI_A\}\le P(A) (E f^2)^{1/2}$ and deduce from
(4.20) with the help of (4.21) the following estimate;
$$
\check Z^{(3)}\le 3 C_0^2 s^3\cdot  \exp\left\{ C_0\mu + 36 V_{12}
\mu \right\} \cdot \left(B(12h \mu^{1/2})\right)^{1/2}\cdot
{(2s)!\over s!\, (s+1)!}
$$
$$
\times  \sum_{d\ge s^{1/2 -\epsilon} }\ \sum_{{K\ge 0}} \ {1\over
h^{K} } \cdot   \exp\left\{ - {C_2\, d\over 6+K}\right\}\cdot
\exp\left\{ {2sd\over n} V_4\right\}.
 \eqno
(4.22)
$$

We see that the problem of the estimate of $\check Z^{(3)}_{2s}$
is reduced to the question about the maximum value of the
expression
$$
F(K)= {2sd\over n} V_4 - {C_2\, d\over 6+K} - K \log h \eqno
$$
as a function of variable $K$.
 Function
$$
f^{(h)}(x) = {2\mu^{1/3} d V_4\over n^{1/3}} - {C_2\, d\over 6+ x}
- x\log h, \quad x\ge  0
$$
takes its maximum value at the point $x_0 = \sqrt{{C_2\, d\over
\log h }}-6$. This gives the estimate
$$
f^{(h)}(x) \le 2 \sqrt d \left({\mu^{1/3} d^{1/2} V_4 \over
n^{1/3}} - \sqrt{C_2\log h}\right) +6 \log h.
$$
Remembering that $s^{1/2-\epsilon}\le d\le s$, we see that if
$
h =h_0 =  \exp\left\{{2+2\mu V_4^2 \over C_2}\right\},
$
 then
$$
F(K)\le f^{(h_0)}_{\max} = -2
\left(n^{2/3}\mu^{1/3}\right)^{1/4-\epsilon/2} +6 \log h_0.
$$

Returning to (4.22), we conclude that  the choice of sufficiently
small positive $\epsilon$, say $\epsilon < 1/6$, leads to the
estimate
$$
\check Z^{(3)}_{2s}\le {3C_0  s^5}\cdot  {(2s)!\over s!\,
(s+1)!}\cdot \exp\{- \mu^{1/24} n^{1/9} + 6\log h_0 +
(C_0+36V_{12})\mu\} \cdot \left(B(12h_0\mu^{1/2})+1\right).
$$
Then obviously $\check Z^{(3)}_{2s} = o(1)$ in the limit
$(s,n)_\mu\to\infty$.

\subsubsection{Estimate of $\tilde Z^{(3)}_{2s}$}

In the present subsection we consider (4.18) with  $\s_4=0 $ and
$r_2 +r_3=R$ denoted by $\tilde Z^{(3)}_{2s}$. To estimate this
sub-sum,  we use the same principle of the estimates as in the
previous subsection: in a part of terms the infinitely increasing
factor $X$ is replaced either by $\Phi= 6sM_\theta/n$ or by $\Psi
= 72V_{12} s^2M_\theta/n^2$ that are asymptotically bounded. All
that we need here is the following elementary computation, where
we use (4.22);
$$
\sum_{\stackrel{r_2+r_3=R}{r_2\le \nu_2, \, r_3\le 2\nu_3}}\ \
  {\tilde X^{\nu_2-r_2}\over (\nu_2
-r_2 )!} \cdot {\Phi^{r_2}\over r_2!} \cdot {\Psi^{\langle
r_3/2\rangle}\over \langle r_3/2\rangle!} \cdot {Y^{\nu_3 -\langle
r_3/2\rangle}\over (\nu_3 - \langle r_3/2\rangle)!}
$$
$$
\le  {\left(\tilde X+h\Phi\right)^{\nu_2}\over \nu_2!}
\sum_{r_3=0}^{2\nu_3} {1\over h^{R-r_3}}\cdot {\Psi^{\langle
r_3/2\rangle}\over \langle r_3/2\rangle!} \cdot {Y^{\nu_3 -\langle
r_3/2\rangle}\over (\nu_3 - \langle r_3/2\rangle)!}
 $$
  $$
 \le
{2\over h^{R}} \cdot {\left(X+h\Phi\right)^{\nu_2}\over \nu_2!}
\cdot {\left(Y+h^2\Psi\right)^{\nu_3}\over \nu_3!} .
 \eqno (4.23)
$$

Regarding (4.18) and using denotation $Y=36V_{12}s^3/n^2$, we get
the following estimate;
$$
\tilde Z^{(3)}_{2s} \le   {n(2s)!\over s!\, (s+1)!} \exp\left\{
C_0 {s^3\over n^2} - {s^2\over 2n}\right\} \ \sum_{\theta\in
\Theta_{2s}} \  \sum_{d\ge s^{1/2-\epsilon}}\ \ \sum_{N=1}^s \ \
\sum_{K\ge 0}\ \
 \ I_ {\Theta^{(d/(2N+K))}_{2s}}(\theta)
 $$
 $$
 \times  \sum_{\s = 0}^{C_0 s^2/n} \ \
 \sum_{\nu_2+2\nu_3=\s}\ \  \sum_{R\ge K} \ \ \sum_{{r_2+
r_3=R}}\  \ \ {1\over r_2!}\cdot \Phi^{r_2}
$$
$$
\times {1\over (\nu_2-r_2)!} \left( X + {2sdV_4\over n}
\right)^{\nu_2-r_2} \cdot {Y^{\nu_3 - \langle r_3/2\rangle}\over
(\nu_3-\langle r_3/2\rangle )!}  \cdot {\Psi^{\langle
r_3/2\rangle}\over \langle r_3/ 2\rangle!}\, .
 \eqno (4.24)
$$
Applying (4.23) with $\tilde X = X + 2sdV_4/n$ to the last two
sums of (4.24) and repeating computations of the previous
subsection, we can write that
$$
\tilde Z^{(3)}_{2s}\le {6C_0  s^5}\cdot \exp\{- \mu^{1/24} n^{1/9}
+ 6\log h_0 + (C_0+36V_{12})\mu\} \cdot
\left(B(24h_0\mu^{1/2})+1\right).
 \eqno (4.25)
$$
Here we have used the fact that $\Psi =o(\Phi)$ as
$(s,n)_\mu\to\infty$. Clearly, $\tilde Z^{(3)}_{2s} = o( 1)$ in
the limit $(s,n)_\mu\to\infty$.

\subsubsection{Estimate of $\breve  Z^{(3)}_{2s}$}

In this subsection we obtain the estimate of the right-hand side
of  (4.18) in the case of $\s_4\ge 1$. It differs from the
previous case of $\s_4=0$ by the factor
$((k-1)\nu_k)^{\rho_k}/\rho_k!$ that estimates the number of
choices of $\rho_4$ instants among $\s_4$ ones. Relation (4.19)
shows that this factor is compensated because of the presence of
the vertices with $\kappa\ge 4$ that produce the factor
$n^{-\s_4/12}$ of (4.19b).

Repeating the arguments and the computations of the previous
subsection, we get the estimate
$$
\breve Z^{(3)}_{2s}\le n \sum_{d=s^{1/2-\epsilon}}^s\
\sum_{\s=0}^{C_0s^2/n}\  \sum_{N=1}^s \ \sum_{K=0}^\s \ \sum_{R\ge
K} \ \ \sum_{r_2+r_3+\rho_4=R}\ \ \sum_{\bar \nu: \, \s_4\ge 1
}^{(N)}\  \ \ { (2s)!\over s!(s+1)!}
$$
$$
\times
 {1\over(\nu_2-r_2)!}
  \left(X+ {2sdV_4\over n}\right)^{\nu_2-r_2}\cdot
 {\Phi^{r_2}\over r_2!} \cdot
{Y^{\nu_3-\langle r_3/2\rangle}\over(\nu_3-\langle
r_3/2\rangle)!}\cdot {\Psi^{\langle r_3/2\rangle}\over \langle
r_3/2\rangle!}
 $$
  $$
\times {1\over n^{\s_4/12}}\cdot {\s_4^{\rho_4}\over \rho_4!}\
\prod_{k=4}^s\ {\d_n^{\nu_k}\over\nu_k!}\cdot H_n^{(k-4)\nu_k}
$$
$$
\times
 \exp\left\{C_0\mu
-{s^2\over 2n}\right\} \cdot \exp\left\{- {C_2d\over 2
N'+K}\right\}\cdot \left({
 U^{2N}C_1^{N}s^{N}\over n^{N-1}}\right)^{I_4(N)},
 \eqno (4.26)
$$
where we denoted $N'=\max\{3,N\}$ and
 $I_4(N) = I_{[4,+\infty)}(N)$.

Let us describe the operations we perform to estimate the
right-hand side of (4.24). First we estimate the sum over all
possible sets $(\nu_4,\nu_5, \dots, \nu_s)$ as follows;
$$
 \sum_{\nu_4+\dots+\nu_s=\s_4\ge 1}\ \ \prod_{k=4}^s\
{\d_n^{\nu_k}\over\nu_k!}\cdot H_n^{(k-4)\nu_k} \le
\exp\left\{\d_n \sum_{k\ge 4}H_n^{k-4}\right\},
 \eqno (4.27)
 $$
where the series over $k$ is obviously convergent. The last
expression tends to $1$ as $(s,n)_\mu\to\infty$.

Next, using an analog of (4.23), we can write that
$$
\sum_{r_2+r_3 = R-\rho_4}  \ {\Phi^{r_2}\over r_2!}
 \cdot{\tilde X
^{\nu_2-r_2}\over(\nu_2-r_2)!}
  \cdot {Y^{\nu_3-\langle r_3/2\rangle}\over
  (\nu_3-\langle r_3/2\rangle)!}\cdot
{\Psi^{\langle r_3/2\rangle}\over \langle r_3/2\rangle!}\
$$
$$
\le {2\over h^{R-\rho_4}\, }\cdot {1\over \nu_2!} (\tilde
X+h\Phi)^{\nu_2} \cdot {1\over \nu_3!} (Y+h^2\Psi)^{\nu_3}.
 \eqno (4.28)
 $$
Finally, we observe that the following inequality is true;
$$
\sum_{R\ge K} {1\over h^R} \sum_{\rho_4=0}^R
{(h\s_4)^{\rho_4}\over \rho_4!} \cdot {1\over n^{\s_4/12}}\le
{h\over h^K(h-1)}\cdot {e^{h \s_4}\over n^{\s_4/12}}.
 \eqno (4.29)
$$
Taking into account (4.27), (4.28), and (4.29), we  derive from
(4.26) inequality
$$
\breve Z^{(3)}_{2s}\le \sum_{d=s^{1/2-\epsilon}}^s\
 \ {n (2s)!\over s!\, (s+1)!}
 \exp\{C_0 \mu + 36V_{12}\mu^3\}\cdot
 \left(B( 24h\mu^{1/2})+1\right)
$$
$$
\times {4C_0 hs^5\over h-1}\  \sum_{N\ge 1} \ \sum_{K\ge 0}\
{1\over h^K}\exp\left\{ {2sdV_4\over n} -
 {C_2d\over 2N'+K}\right\} \cdot
  \left( {U^{2N}C_1^{N}s^{N}\over n^{N-1}}\right)^{I_4(N)}
 \eqno (4.30)
$$
for all $n$ such that for $n\ge \exp\{12h\}$.

 Similarly to the
situation encountered in (4.22), we consider
 the following function of two variables
$$
F(N,K) = {2sdV_4\over n} - {C_2d\over 2\max(3,N)+K} - K\log h -
{N-3\over 3}I_4(N)  \log n.
$$
Denoting $N''= (N-3)I_{[4,+\infty)}(N)$, we see that the problem
of estimate of $\breve Z^{(3)}_{2s}$ for large enough values of
$n$ such that $\log n^{1/6}\ge \log h$ is reduced to the study of
the maximal value of the function
$$
\breve F(N'',K) = {2sd V_4\over n} - {C_2 d\over 6 + K + 2N''} -
(K+2N'')\log h.
$$
This maximum corresponds to the value $f^{(h_0)}_{\max}$
determined in the previous subsections and we can write that
$$
F(N,K) \le \breve F(N'',K) \le
 -2
\left(n^{2/3}\mu^{1/3}\right)^{1/4-\epsilon/2} +6 \log h_0
$$
provided $\log n\ge 6 \log h_0$.

Then we deduce from (4.30) inequality
$$
\breve Z^{(3)}_{2s}\le { n(2s)!\over s!\, (s+1)!} \cdot
 \exp\{C_0 \mu + 36V_{12}\mu^3\}\cdot
 \left(B( 24h\mu^{1/2})+1\right)
 $$
 $$
 \times{4C_0
h_0\, s^7\over h_0-1}\cdot \exp \left\{  -2
\left(n^{1/6-\epsilon/3}\mu^{1/12-\epsilon/6}\right) +6 \log
h_0\right\}
 \eqno (4.31)
$$
that holds for all $n$ such that $n\ge \exp\{ 12h_0\}$. We see
that the choice of $0<\epsilon<1/6$ makes the product of the last
two terms of (4.28) vanishing in the limit $(s,n)_\mu\to\infty$.
Then $\breve Z^{(3)}_{2s}=o(1)$ as $(s,n)_\mu\to\infty$ provided
(4.1) holds.

This result together with (4.25) shows that the estimate of
$Z^{(3)}_{2s}$ is completed.


\subsection{Estimate of $Z^{(4)}_{2s}$}

In this part, we follow the general description of the walks, and
give the estimate of $Z^{(4)}_{2s}$ in the way that slightly
differs from that presented in \cite{R}.

Remembering (3.10) and $W_n$, we can write that
$$
Z^{(4)}_{2s}\le{ (2s)!\over  s!\, (s+1)!}\ \sum_{\s\ge {C_0s^2/
n}} \ \  \sum_{\bar \nu: \, \vert \bar \nu\vert_1 =\s} \
{n(n-1)\cdots (n-s+\s)\over n^s}\,
  \prod_{k=2}^s\left({ (2k U^2 s)^k\over k!}\right)^{\nu_k}.
$$

The main difference between this sub-sum and the previous ones is
that in (4.8) the factor $\exp\{s\s\}$ is not bounded for $\s\ge
C_0 s^2/n$ and $\exp\{-s^2/n\}$ cannot be used as the normalizing
factor for the terms $(s^2/n)^\s /\s!$. In this case the last
expression is bounded by itself. This is the main idea of the
proof of the estimate of $Z^{(4)}_{2s}$ proposed in \cite{R}. We
reconstruct this proof with slight modifications and corrections.

Denoting
$$
 \vert \bar \nu\vert_2 = \sum_{k\ge 2}(k-2)\nu_k,
$$
we can write that
$$
\prod_{k=2}^s s^{k\nu_k} = s^{2\s - \chi}.
$$
Then
$$
Z^{(4)}_{2s}\le{n (2s)!\over \s!\, (s+1)!}\ \sum_{\s\ge C_0s^2/n}
\ \  \sum_{\chi=0}^\s\ \ \ \sum_{\bar \nu: \, \vert \bar
\nu\vert_1 =\s, \vert \bar\nu \vert_2=\chi} \
{(n-1)\cdots(n-s+\s)\over n^{s-\s}}\cdot
 {1\over n^\s}
$$
$$
\times \ {s^{2\s}\over s^\chi}\cdot {1\over \nu_2!\,
\nu_3!\,\cdots \nu_s!}\ \prod_{k=2}^s \left(C_1
U^2\right)^{k\nu_k}.
$$
\V \noindent
 Multiplying and dividing by $\s!$ and by
$(\s-\chi)!$, we obtain inequality
 \v
$$
Z^{(4)}_{2s}\le{n (2s)!\over \s!\, (s+1)!} \ \sum_{\s\ge C_0s^2/n}
\ \
 {s^{2\s}\over \s!\, n^\s} \ \  \sum_{\chi=0}^\s
   \  {\s!\over s^\chi \
(\s-\chi)!}
$$
$$
\times  \sum_{\bar \nu: \, \vert \bar \nu\vert_1 =\s,
\vert\bar\nu\vert_2=\chi} {(\s-\chi)! \over \nu_2!\ \nu_3!\,\cdots
\nu_s!}\ \prod_{k=2}^s \left(C_1 U^2\right)^{k\nu_k}.
$$
\V \noindent Using the definition of $\vert\bar\nu\vert_2$, we can
rewrite the previous estimate in the form
 \v
 $$
Z^{(4)}_{2s}\le{n (2s)!\over s!\, (s+1)!}\  \ \sum_{\s\ge
C_0s^2/n} \ \sum_{\chi=0}^\s \ {1\over \s!} \left({s^2\over
n}\right)^{\s}
$$
$$
\times
 \sum_{\bar \nu: \, \vert \bar \nu\vert_1 =\s, \,\vert \bar \nu\vert_2=\chi}
\ {(\s-\chi)! \over \nu_2!\ \nu_3!\,\cdots \nu_s!}\ \prod_{k=2}^s
\left({C_1 U^2\s\over s}\right)^{(k-2)\nu_k}
 \left(C_1 U^2\right)^{2\nu_k}.
 \eqno (4. 32)
$$

\V \noindent
 At this point we separate the sum over $\s$ into two
parts: the first sub-sum we denote by $\dot Z^{(4)}_{2s}$
corresponds to the interval  ${\displaystyle
C_0s^2\over\displaystyle  n}\le \s\le {\displaystyle s\over
\displaystyle  2C_1U^2}$, the second denoted by $\ddot
Z^{(4)}_{2s}$ corresponds to the remaining part
 $\s\ge {\displaystyle s\over \displaystyle  2C_1U^2}$.

 Aiming the estimate of $\dot Z^{(4)}_{2s}$, we use
the multinomial theorem in the form
 \v
$$
\sum_{\nu_2+\dots+\nu_s = \s-\chi} {(\s-\chi)!\over \nu_2!\cdots
\nu_s!} \ \prod_{k=2}^s{1\over 2^{(k-2)\nu_k}} \le
 \left(1+{1\over 2}+\dots+{1\over
2^{s-2}}\right)^{\s-\chi}\le 2^{\s-\chi}.
$$
\V
 \noindent
  Since  \mbox{$2C_1^2U^4>1$,} we deduce from (4.32) with the
help of the Stirling formula that
 \newpage
$$
\dot Z^{(4)}_{2s}\le{n (2s)!\over s!\, (s+1)!}\ \ \sum_{
C_0s^2/n\le \s \le s/(2C_1U^2)} \ \ {1\over \s!}\left({s^2\over
n}\right)^{\s}\ \sum_{\chi=0}^\s (2C_1^2U^4)^{\s-\chi}
$$
$$
 \le {n (2s)!\over s!\, (s+1)!}\ \ \sum_{ \s\ge C_0s^2/n} \
{(2C_1^2U^4)^{\s+1}\over 2C_1^2U^4-1}\cdot \left({e s^2\over
n\s}\right)^{\s}\cdot {1+o(1)\over \sqrt{2\pi\s}}
$$
$$
 \le{n (2s)!\over s!\, (s+1)!}\cdot {2C_1^2U^4\over 2C_1^2U^4-1}\ \
\sum_{ \s\ge C_0s^2/n} \ \left({2C_1^2U^4 e \over
C_0}\right)^{\s}\cdot {2\over \sqrt{\pi\s}}
$$
$$
 \le {n (2s)!\over s!\ (s+1)!} \cdot {2C_1^2U^4\over 2C_1^2U^4-1}\cdot
{2\over \sqrt{\pi C_0 s^2/n}}\cdot \left( {2C_1^2U^4e\over
C_0}\right)^{C_0 s^2/n}.
 \eqno (4.33)
$$
The product of two last terms vanishes as $(s,n)_\mu\to\infty$ in
the case when
 $
C_0\ge 2eC_1^2U^4.
$

\v  Now it remains to consider the part that is complementary to
$\dot Z^{(4)}_{2s}$; it is estimated by the following expression
 $$
  \ddot Z^{(4)}_{2s}\le {n (2s)!\over s!\,
(s+1)!} \ \sum_{\s\ge s/(2C_1U^2)}\ \ \sum_{\chi=0}^\s \  \left(
{s^2\over n}\right)^{\s}
 \ {1\over s^\chi}
 $$
 $$
 \times\
\sum_{\bar \nu:\, \vert \bar \nu\vert_1=\s, \vert\bar
\nu\vert_2=\chi}
 {1\over \nu_2!\cdots \nu_s!}\ \ \prod_{k=2}^s (C_1
U^2)^{k\nu_k}.
 \eqno (4.34)
$$
Using the identity $\prod_{k\ge 2} (2C_1U^2)^{(k-1)\nu_k} =
(2C_1U^2)^\s$, we can replace the product of the last four factors
of (4.34)  by
$$
{1\over \s!}\cdot\left( {2C_1U^2s^2\over n}\right)^\s\cdot
{\s^\chi\over s^\chi}\cdot {\s!\over (\s-\chi)! \  \s^\chi}\cdot
{(\s-\chi)!\over \nu_2!\cdots \nu_s!}\ \  \prod_{k= 2}^s
\left({C_1U^2\over 2^{k-1}}\right)^{\nu_k}.
$$
Using again the Stirling formula, remembering that $\s\le s$, and
applying the  multinomial theorem to the sum over $\bar \nu$, we
get  that
$$
\ddot Z^{(4)}_{2s}\le {n (2s)!\over s!\, (s+1)!} \ \sum_{\s\ge
s/(2C_1U^2)}\ \sqrt{{\s\over 2\pi}}\cdot  \left(
{4C_1^3U^6\mu^{1/3} e\over n^{1/3}}\right)^{\s}.
$$
The last series is obviously $o(1)$ as $(s,n)_\mu \to\infty$. The
estimate of $Z^{(4)}_{2s}$ is completed.

\newpage

\section{More examples of the walks}

Let us consider a walk $ W^{(0)}_{18}$ that has a number of open
simple self-intersections.

\begin{figure}[htpb]
\centerline{\includegraphics[width=12cm]{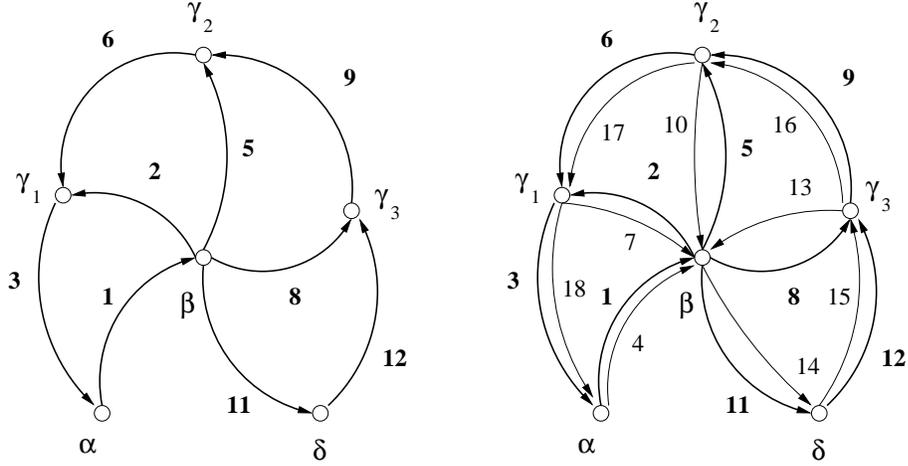}}
\caption{\footnotesize{The marked edges of the graph
$g(W^{(0)}_{18})$ and the full graph $g(W_{18}^{(0)})$}}
\end{figure}

Basing on the example given on Figure 4, one can easily construct
a sequence of walks of $ 2s$ steps  such that their
 graphs contain a vertex with the exit
degree that infinitely increases as $s\to\infty$; the
self-intersection degree of this vertex remains bounded and the
corresponding Dyck path $\theta_{2s}$ is such that its tree
$\CT(\theta_{2s})$ has no vertices with large degree. Below we
present one of the possible examples.

Regarding the  walk $W^{(0)}_{2s}$ with $2s=18$ steps depicted on
Figure 4, we see that the marked instants are given by the
instants of time $1,2,3,5,6,8,9,11,12$ and  the vertex $\b$ is the
vertex of the self-intersection degree 1. This walk contains four
BTS-instants given by $3,5,8,11$. There are four imported cells at
the vertex $\b$ determined by the instants of time $4,7,10$ and
$13$.

Let us modify this walk $W^{(0)}$ to $W^{(1)}$ by adding some
"there-and-back" steps after each non-marked arrival $(\g_i,\b)$,
$ i=1,2,3$; let them be three each time. Then the graph of the
walk is added by nine edges and  we get the walk
$$
W^{(1)} = (\a,\b,\g_1,\a,\b,\g_2,\g_1,\b,
\underbrace{\varepsilon_1,\b,\eta_1,\b,\mu_1,\b}_{{\hbox{three
edges}}},\g_3 ,\g_2,\b,
\underbrace{\varepsilon_2,\b,\eta_2,\b,\mu_2,\b}_{\hbox{three
edges}}, \d,\g_3,...\}
$$
The number of marked steps that leave $\b$ denoted by $\nu_n$ in
papers
 \cite{S} and  \cite{R} and by $\deg_e(\b)$ in the present paper  is
increased by $9$ but the self-intersection degree of $\b$ is still
equal to one;  $\kappa(\b)=N=1$.

Certainly, one can consider analogs of $W^{(1)}$ with  more
vertices of the type $\g_i$, say ten:  $\g_1,..., \g_{Q}$, $Q=10$.
If we add ten triplets of "there-and-back" edges  with vertices
$\varepsilon_i,\eta_i,\mu_i$ and pass them  after each arrival to
$\b$ by non-marked steps $(\g_i,\g)$, respectively, then we get a
walk $W^{(2)}(Q)$ with $\deg_e(\b)=\nu_n = 3Q +Q+1=41$ and still
$\kappa(\b)=1$. If $Q\to\infty$, then $\deg_e(\b)$ infinitely
increases.

Regarding the walks $W^{(2)}(Q)$ with arbitrarily large $Q$, we
see that the expression $\exp\{c\deg_e(\b) s_n/n\}$ of \mbox{(4.14
\cite{R})} goes out of the control in the limit of large $s$
because it cannot be suppressed by the factor of the form
$s^N/n^{N-1}$ with $N=\kappa(\b)$, $\kappa(\b)=1$.

From another hand, the trees $\CT(W^{(2)}(Q))$ are such that
 for any value of $Q$ there is no  vertices in
$\CT(W^{(2)}(Q))$ with the exit degree greater than $5$. This
situation is possible because the groups of three edges are
imported at $\b$ from different parts of the tree $\CT$; freely
expressed, these edges grow in $g(W^{(2)}(Q))$ from imported
cells. As we have seen in Section 2, the presence   of the
imported cells is possible due to the presence of BTS-instants in
the walk. Thus we conclude that the trees under consideration do
not verify the $\CL(6)$-property (see subsection 4.1) and
therefore there is no factors with the exponential estimates of
the form (4.17) that would suppress the growth of \mbox{(4.14
\cite{R})}. Therefore the proof of the estimate of $Z^{(3)}_{2s}$
presented in \cite{R} is not correct. The same is true with
respect to the proof of the Lemma 3 of \cite {SS2}.

\section{Appendix}

The main aim of the paper \cite{R} was to extend the universality
results of the papers \cite{SS2,S} to the more general case when
the entries of the Wigner random matrices $a_{ij}$ (1.1), (1.2)
are given by random variables with polynomially decaying
probability distribution. We are going to prove the statement that
shows that this is indeed the case. To do this, we need just a
slight modification of the proof of Theorem 4.1.

In paper \cite{R} the Wigner ensemble  is considered, where the
random variables $a_{ij}$ verify  condition $P\{\vert a_{ij}\vert
>x\}\le x^{-18}$. However, to make  inequality
(4.7 \cite{R}) true, one has to require more restrictive
conditions, say with exponent $18$ replaced by $36$. In the
present paper, we do not aim the optimal conditions for $a_{ij}$
and prove our statement in the frameworks of \cite{R} and
\cite{S}; modifications of the proof of Theorem 4.1 necessitate
also more conditions on $a_{ij}$ than those of \cite{R} in the
case of polynomially bounded random variables.

\v {\bf Theorem 6.1} {\it Let us consider the random matrix
ensemble described in Theorem 4.1 with the bound (4.2) replaced by
the following condition;
$$
\E \vert a_{ij}\vert ^q<\infty
$$
for all $q\le q_0 = 76$. Then the estimate (4.4) is true in the
limit $(s,n)_\mu\to\infty$, where the constants do not depend on
particular values of the moments $\E a_{ij}^{2k}$, $k\ge 2$. }

\v {\it Proof.} By the standard approach of the probability
theory, we introduce  the truncated random variables
$$
\hat a_{ij} = \cases{
 a_{ij},& if $ \vert a_{ij}\vert \le U_n$;\cr
 0,& if $ \vert a_{ij}\vert > U_n$,\cr}
\eqno (6.2)
$$
where $U_n = n^{\a}$ with $\a=1/25$. Then we consider the random
matrices $\hat A_{ij} = \hat a_{ij}/n^{1/2}$ and write down
equality
 $$
 \E \left\{ \T \hat A_n^{2s} \right\}= \sum_{l=1}^4 \hat
 Z^{(l)}_{2s},
 $$
 where  $\hat Z^{(l)}_{2s}$ are determined exactly as it is done in (4.5).
We are going to show that these sub-sums admit the same asymptotic
estimates as $Z^{(l)}_{2s}$ of (4.5).

The first sub-sum $\hat Z^{(1)}_{2s}$ is estimated as
$Z^{(1)}_{2s}$ (4.11) with no changes.

To estimate $Z^{(2)}_{2s}$, we repeat the reasonings of subsection
4.3.1 that lead to inequality (4.14) and (4.15). The estimate
$Z^{(2)}_{2s} = o(1)$ as $(s,n)_\mu\to\infty$ is valid due to
relations
$$
{C_1^k U_n^{2k} s^k\over n^{k-1}} = C_1^k \mu ^{k/3} {n^{2k/3 +
2k/25}\over n^{k-1}} = C_1^k\mu^{k/3} n^{1-19k/75}\to 0, \quad
n\to\infty
 \eqno (6.3)
$$
for all $k\ge 4$.

 To estimate $\hat Z^{(3)}_{2s}$, we introduce a
slight modification of the computations  used to estimate
$Z^{(3)}_{2s}$. All that we need here is to redefine the variables
$\d_n$ and $H_n$ of (4.19). We rewrite (4.19a) in the form
$$
C_1^k U_n^{2k}\
 {n^{2/3}\over n^{(k-1)/3}} = C_1^4 \mu ^{4/3}\
 {n ^{8\a +2/3}\over n^{1-\b}}
  \left( {C_1 \mu^{1/3}
 n^{2\a}\over n^{(1-\b)/3}}\right)^{k-4}
  {1\over n^{(k-1)\b/3}} = \hat \d_n \hat H_n^{k-4} {1\over n^{(k-1)\b/3}}.
 \eqno (6.4)
 $$
If $8\a+ {2\over 3} \le 1 - \b$, then $\hat \d_n = O(1)$ and $\hat
H_n\to 0$.
 The choice of $\b=1/75$ leads to the value $\a=1/25$
imposed in (6.2). We see that the value $\a_0= 1/24$ represents
the lower bound for $\a$ in the approach developed.

Then we can use the analogue of (4.19b) with $n^{\s_4/12}$
replaced by $n^{\s_4/225}$. All other computations that lead to
the estimate of $Z^{(3)}_{2s}$ can be repeated as they are.

The estimate of $\hat Z^{(4)}_{2s}$ requires somehow more work. To
estimate this sub-sum, let us prove the following auxiliary
statement.

\v {\bf Lemma 6.1} {\it Given any walk $w_{2s}$ of the type $\bar
\nu$, the weight $\CQ (w_{2s}) $ (2.1) is bounded as follows;
 $$
 \CQ(w_{2s}) \le  \prod_{k=2}^s \left( V_{12}\,  U_n^{2(k-2)}
 \right)^{\nu_k}.
 \eqno (6.4)
  $$}

\v {\it Proof.} Let us consider a vertex $\g$ with
 $\kappa(\g)\ge 2 $ of the multi-graph
  $g(w_{2s}) = (\CV,\CE)$
 and color in certain color the first two marked
arrival edges at $\g$ and their non-marked closures. Passing to
another vertex with $\kappa\ge 2$, we repeat the same procedure
and finally get $4\sum_{k=2}^s\nu_k$ colored edges. Clearly, it
remains $2\nu_1 + 2\sum_{k=2}^s (k-2)\nu_k$ non-colored (grey)
edges in $\CE(g)$. Let us remove from the graph $(\CV,\CE)$ all
grey edges excepting the marked edges $e'_j$ whose heads are the
vertices of $\CN_1$; also we  do not remove the closures of these
edges $e'_j$. We denote the remaining graph by $g^\circ=
(\CV,\CE^\circ)$.  The number of the edges removed from $g$ is
greater or equal to $2\sum_{k=2}^s (k-2)\nu_k$. When estimating
the weight of $w_{2s}$, we replace corresponding random variables
by non-random bounds $U_n$ when the removed marked edges end at
the vertices with $\kappa\ge 2$. Then we can write that
$$
\CQ(w_{2s}) \le U_n^{2(k-2)\nu_k}\cdot  \CQ^\circ (w_{2s}),
$$
where $\CQ^\circ(w_{2s})$ represents the product of the
mathematical expectations of the random variables associated with
the edges of $\CE^\circ$. Obviously, we did not replace by $U_n$
those random variables that give factors $V_2=1$.

In the remaining (possibly non-connected) graph $g^\circ$  the set
$\CV$ contains a subset $\CV^\circ$  such that if $\b\in
\CV^\circ$, then $\b$ is the head of of two colored marked edges.
Clearly, $\vert \CV^\circ\vert = \sum_{k=2}^s \nu_k$.
 Regarding a
head $\b \in \CV^\circ $ of two marked colored edges, we see that
their tails $\a'$ and $\a''$ can be either equal or distinct.

Let us consider first the case when the tails are distinct,
 $\a'\neq \a''$.  Then each of the multi-edges
 $(\a',\b)$ and $(\a'',\b)$ can produce factors $V_4 $ or $V_6$ in
 dependence how many edges arrive at $\a'$ and $\a''$ from $\b$.
 Then the contribution of the vertex $\b$ to $\CQ^\circ(w_{2s})$ is
 bounded by $V_6^2\le V_{12}$.

Now let us consider the case when the tails are equal,
$\a'=\a''=\a$. Then the multi-edge $\vert \a,\b\vert$ produces the
factors equal to either $V_4$ or $V_6$ or $V_8$ in dependence of
how many marked edges of $\CE^\circ$ arrive at $\a$ from $\b$.
This can be either one grey edge or two colored edges. Since
 $1\le V_4\le V_6\le V_8$, then the contribution of this vertex $\b$ to
$\CQ^\circ(w_{2s})$ is bounded by $V_8\le V_{12}$. Here we have
used inequality $\kappa(\a)\ge 1,\ \a\in\CV(g_{2s})$.

Collecting the contributions of all vertices of $\CV^\circ$, we
get the estimate
 $$
 \CQ^\circ(w_{2s})\le \prod_{k=2}^s V_{12}^{ \nu_k}.
 $$
It is clear that we do not need to consider those vertices of
$\CV\setminus\CV^\circ$ that are the heads of non-marked edges
because the contributions of the corresponding random variables is
already taken into account. Lemma 6.1 is proved. $\diamond$

\v  Now it is easy to complete the estimate of $\hat
Z^{(4)}_{2s}$. Indeed, formula (4.32) can be rewritten in the form
\v
$$
\hat Z^{(4)}_{2s}\le{n (2s)!\over s!\, (s+1)!}\
  \ \sum_{\s\ge C_0s^2/n} \
   \  \sum_{\chi=0}^\s \ {1\over \s!} \left({s^2\over n}\right)^{\s}\ \
$$
$$
\times\
 \sum_{\bar \nu: \, \vert \bar \nu\vert_1 =\s, \,\vert \bar \nu\vert_2=\chi}
\ {(\s-\chi)! \over \nu_2!\ \nu_3!\,\cdots \nu_s!}\ \prod_{k=2}^s
\left({C_1 U^2_n\s\over s}\right)^{(k-2)\nu_k}
 \left(C_1^2 V_{12}\right)^{\nu_k}.
 \eqno (6.5)
$$
Let us consider first the interval ${C_0s^2/n}\le \s\le
{s/(2C_1U^2_n)}$. Using the multinomial theorem and the Stirling
formula, we get from (6.5) the following analog of the estimate
(4.32);
$$
{\dot {\hat Z}}^{(4)}_{2s} \le{n (2s)!\over s!\, (s+1)!}\ \ \sum_{
C_0s^2/n\le \s \le s/(2C_1U^2_n)} \ \ {1\over \s!}\left({s^2\over
n}\right)^{\s}\ \sum_{\chi=0}^\s (2C_1^2V_{12})^{\s-\chi}
$$
$$
 \le {n (2s)!\over s!\ (s+1)!} \cdot {2C_1^2V_{12}\over 2C_1^2V_{12}-1}\cdot
{2\over \sqrt{\pi C_0 s^2/n}}\cdot \left( {2eC_1^2V_{12}\over
C_0}\right)^{C_0 s^2/n}.
$$
This expression is $o(1)$ in the limit $(s,n)_\mu\to\infty$
provided $C_0\ge 2eC_1^2V_{12}$.

To estimate the sub-sum ${\ddot{\hat Z}}^{(4)}_{2s}$ that
corresponds to the interval $\s\ge s/(2C_1U_n^2)$, we repeat word
by word the computations presented at the end of Section 4 (see
formulas (4.34) and (4.35)).

Summing up the arguments presented above, we see that the
following analog of (4.4)
$$
{\sqrt{\pi \mu}\over 4^s} \E\left\{ \T \hat A_n^{2s}\right\}\le
B(6\mu^{1/2}) \exp\{(36+C_0)\mu\}
 \eqno(6.6)
$$
is valid in the limit $(s,n)_\mu\to \infty$.

By the standard arguments of the probability theory, we have
$$
P\{A_n \neq \hat A_n\}  \le 1-  \left(1- n^{-q_0\a} \E \vert
a_{ij}\vert^{q_0}\right)^{n(n+1)/2} = O(n^{-1-\delta}),\quad
\delta
>0
 $$
 as $n\to \infty$.
Then by the Borel-Cantelli lemma,
$$
P\{A_n \neq \hat A_n \ \ {\hbox{infinitely often}} \} =0.
 \eqno (6.7)
$$
 Relations (6.6) and (6.7) complete the proof of Theorem 6.1.

\end{document}